**Recent selective sweeps in North American *Drosophila melanogaster* show signatures of soft sweeps**

Nandita R. Garud[1,2], Philipp W. Messer[2], Erkan O. Buzbas[2,3], and Dmitri A. Petrov[2]

1. Department of Genetics, Stanford University, Stanford, CA 94305, USA
2. Department of Biology, Stanford University, Stanford, CA 94305, USA
3. Department of Statistical Science, University of Idaho, Moscow, ID 83844, USA

Corresponding Authors:
Nandita R. Garud (ngarud@stanford.edu) and Dmitri A. Petrov (dpetrov@stanford.edu)
Department of Biology
Stanford University
Stanford, CA 94305-5020

**ABSTRACT**

Rapid adaptation has been observed in numerous organisms in response to selective pressures, such as the application of pesticides and the presence of pathogens. When rapid adaptation is driven by rare alleles from the standing genetic variation or by a high population rate of *de novo* adaptive mutation, positive selection should commonly generate soft rather that hard selective sweeps. In a soft sweep, multiple adaptive haplotypes sweep through the population simultaneously, in contrast to hard sweeps in which only a single adaptive haplotype rises to high frequency. Current statistical methods were not designed to detect soft sweeps, and are therefore likely to miss these possibly numerous adaptive events. Here, we develop a statistical test (H12) based on haplotype homozygosity that is capable of detecting both hard and soft sweeps with similar power. We use H12 to identify multiple genomic regions that have undergone recent and strong adaptation in a population sample of fully sequenced *Drosophila melanogaster* strains from the Drosophila Genetic Reference Panel (DGRP). Visual inspection of the top 50 peaks revealed that multiple haplotypes are at high frequency, consistent with signatures of soft sweep. We developed a second statistic (H2/H1) that is sensitive to signatures common to soft sweeps but not hard sweeps, in order to determine whether sweeps detected by H12 can be more easily generated by hard versus soft sweeps. Surprisingly, we find that the H12 and H2/H1 values for all top 50 peaks are more easily generated by soft sweeps than hard sweeps under several evolutionary scenarios.

## AUTHOR SUMMARY

Evolutionary adaptation is a process in which beneficial mutations increase in frequency in response to selective pressures. If these mutations were previously rare or absent from the population, adaptation should generate a characteristic signature in the genetic diversity around the adaptive locus, known as a selective sweep. Such selective sweeps can be distinguished into *hard* selective sweeps, where only a single adaptive mutation rises in frequency, or *soft* selective sweeps, where multiple adaptive mutations at the same locus sweep through the population simultaneously. Here we design a new statistical method that can identify both hard and soft sweeps in population genomic data and apply this method to a *Drosophila melanogaster* population genomic dataset consisting of 145 sequenced strains collected in North Carolina. We find that selective sweeps were in fact quite abundant in the recent history of this population. Interestingly, we also find that practically all of the strongest and most recent sweeps show patterns that seem more consistent with soft rather than hard sweeps. We discuss the implications of these findings for the discovery and quantification of adaptation from population genomic data.

## INTRODUCTION

The ability to identify the genomic loci subject to recent positive selection is essential for our efforts to uncover the genetic basis of phenotypic evolution and to understand the overall role of adaptation in molecular evolution. The fruit fly *Drosophila melanogaster* is one of the classic model organisms for studying the molecular basis and signatures of adaptation. Recent studies have provided evidence for pervasive molecular adaptation in this species, suggesting that approximately 50% of the amino acid changing substitutions, and similarly large proportions of non-coding substitutions, were adaptive [1,2,3,4,5,6].

Adaptation in *D. melanogaster* is not only common but also, at times, extremely rapid. For example, resistance to the most commonly used pesticides, carbamates and organophosphates, evolved within decades after their introduction primarily through three point mutations at highly conserved sites in the gene *Ace*, which encodes for the neuronal signaling enzyme Acetylcholinesterase [7,8,9]. Similarly, resistance to DDT evolved within a few years primarily *via* the insertion of an *Accord* transposon in the 5' regulatory region of the gene *Cyp6g1* and other complex mutations involving this locus [10,11]. Increased resistance to infection by the sigma virus, as well as resistance to certain organophosphates, has been associated with a transposable element insertion in the protein-coding region of the gene *CHKov1* [12,13]. In all these cases, the adaptive alleles were initially absent in the population (*Ace* and *Cyp6g1*), or present at only a very low frequency (*CHKov1*), but quickly became prevalent in a selective sweep.

Intriguingly, in-depth studies of the population genetic signatures of adaptation at these three loci [9,11,13] revealed that none produced the expected signatures of a classic hard selective sweep, in which a single adaptive haplotype rises in frequency and removes genetic diversity in the vicinity of the adaptive locus [14,15]. Instead, all three cases of adaptation produced signatures of multiple haplotypes bearing adaptive alleles at high frequencies, compatible with a signature of a 'soft' selective sweep [16,17]. A soft sweep can arise when multiple adaptive alleles are already present in the population as standing genetic variation (SGV) at the onset of positive selection or from multiple *de novo* adaptive mutations entering the population simultaneously during a sweep [16,17,18,19,20]. In the cases of *Ace* and *Cyp6g1*, soft sweeps involved multiple *de novo* mutations [9,11,13], whereas in the case of *CHKov1*, a soft sweep arose from SGV [9,11,13].

Unfortunately, most scans for selective sweeps in population genomic data were built around the paradigm of hard selective sweeps (although see [21]) and focus on associated signatures such as a dip in neutral diversity around the selected site [15], an excess of low frequency alleles or high-frequency derived alleles in the frequency spectrum of neutral polymorphisms surrounding the selected site (Tajima's *D*, Fay and Wu's *H*, Sweepfinder) [22,23,24], or the observation of a single long haplotype present at high frequency in the population (*iHS*) [25,26]. In a soft sweep, however, several haplotypes can be frequent in the population at the selected locus and thus neutral diversity will be reduced to a lesser extent than in a hard sweep. As a result, as Pennings *et al.* showed, methods based on the levels and frequency distributions of neutral diversity have low power to detect soft sweeps [20,27]. Given that none of the described cases of rapid adaptation via soft sweeps at *Ace, Cyp6g1,* and *CHKov1* were discovered by genomic scans relying on signatures of adaptation left in patterns of polymorphism, it is possible that additional cases of selective sweeps in *D. melanogaster* remain to be discovered in a systematic scan that can detect signatures of both hard and soft sweeps.

Some genomic signatures do have power to detect hard and soft sweeps. In particular, linkage disequilibrium (LD), both between pairs of sites and measured as haplotype homozygosity, should be elevated over neutral levels in both hard and soft sweeps, as long as the number of independent haplotypes bearing the adaptive allele (the 'softness' of the sweep) is not too high [27,28].

The softness of a sweep should depend on the number of independent haplotypes that rise in frequency simultaneously, with greater softness of the sweeps expected either (i) when the rate of mutation to *de novo* adaptive alleles at a locus is very high and multiple alleles quickly arise and establish after the onset of selection, or (ii) when adaptation uses SGV with previously neutral or deleterious alleles that are present at higher frequency at the onset of selection [16,17].

More specifically, for sweeps arising from multiple *de novo* mutations, Pennings and Hermisson [17] showed that the key population genetic parameter that determines the softness of the sweep is $\theta_A = 4N_e\mu_A$, proportional to the product of $N_e$, the effective population size estimated over the period relevant for adaptation [29,30], and $\mu_A$, the mutation rate toward adaptive alleles at a locus per individual per generation. The mutation-limited regime with hard selective sweeps corresponds to $\theta_A << 1$, whereas $\theta_A > 1$ specifies the non-mutation-limited regime with primarily soft sweeps. As $\theta_A$ becomes larger, the sweeps become softer as more haplotypes increase in frequency simultaneously [17]. In the case of sweeps arising from SGV, the softness of a sweep is governed by the starting partial frequency of the adaptive allele. For any given rate of recombination, adaptive alleles starting at a higher frequency at the onset of selection should be older and should thus be present on more distinct haplotypes and give rise to softer sweeps [16].

Here we develop a statistical test based on modified haplotype homozygosity for detecting selective sweeps in population genomic data that has similar power for identifying both hard and soft sweeps of varying softness. We apply this test in a genome-wide scan in *D. melanogaster* using the Drosophila Genetic Reference Panel (DGRP) data set [31] consisting of 162 fully sequenced isogenic strains from a North Carolina population. Our scan recovers all three previously known examples of soft sweeps at the loci *Ace*, *Cyp6g1*, and *CHKov1*, and additionally identifies pervasive signatures of recent selective sweeps at a large number of previously unknown loci. We examine the haplotype frequency spectra at loci of the top 50 candidate sweeps in our scan and find that multiple haplotypes are at high frequencies in all 50 sweeps. In comparison to expectations under several tested neutral demographic scenarios, we see multiple haplotypes at high frequency with high haplotype homozygosity. Strikingly, we find that the signatures at the top 50 candidate peaks can be most easily generated by soft sweeps from multiple *de novo* origins or low-frequency SGV, and cannot be easily generated by a classical hard sweep or neutral demographic signatures.

## RESULTS

### Haplotype homozygosity statistics to detect hard and soft sweeps in DGRP data

In this section we describe a new haplotype statistic that can be used to detect recent and strong sweeps in the DGRP data. Our statistic is based on the reasoning that linkage disequilibrium (LD), measured between pairs of sites or as haplotype homozygosity, should be a sensitive statistic in the detection of both hard and soft sweeps [27,28] as long as the distance between sites or the length of haplotypes are large enough that demographic processes are unlikely to generate these signatures neutrally. At the same time, the analysis window must not be so large that most adaptive events of reasonable selective strength fail to generate sweeps that span the whole window. To determine an appropriate analysis window size, we compared the decay in LD in DGRP data with that of expectations under several realistic neutral demographic models for North American *D. melanogaster*.

We considered six demographic models for North American *D. melanogaster* for our analysis (Figure 1). The first demographic model is an admixture model of the North American *D. melanogaster* population proposed by Duchen *et al.* [32]. In this model, the North American population was co-founded by flies both from Africa and Europe approximately $3.05\times10^{-4}N_e$ generations ago (where $N_e \approx 5x10^6$). The second model is a modified admixture model, also proposed by Duchen *et al.* [32], in which the founding European population underwent a bottleneck before the admixture event (see Table S1 for complete parameterization of both admixture models). The third model is a constant population model with an effective population size of $N_e = 10^6$ [33,34], which we considered for its simplicity and computational feasibility. In addition, we inferred a constant $N_e = 2.7x10^6$ demographic model fit to Watterson's $\theta_W$ measured in short intron autosomal polymorphism data from the DGRP data set. Finally, we fit several simple out-of-Africa bottleneck models to short intron regions in the DGRP data set using the software DaDi [35] (Table S2) (Methods). The two bottleneck models we ultimately used are a severe but short bottleneck model ($N_B$=0.002, $T_B$=0.0002) and a shallow but long bottleneck model ($N_B$=0.4, $T_B$=0.0560), both of which fit the data equally well among a range of other inferred bottleneck models (see Figure S1 for parameterization). All models except for the constant $N_e = 10^6$ model fit the DGRP short intron data in terms of site frequency spectrum (SFS) summary statistics, such as the number of segregating sites ($S$) and pair wise diversity ($\pi$) (Table S3).

We compared the decay in pair-wise LD in the DGRP data at distances from a few base pairs to 10Kb with the expectations under each of the six demographic models (Figure 2). In our simulations of neutral demographic scenarios, we used parameters relevant for our subsequent analysis of the DGRP data. To match the sample depth of the DGRP data set (145 strains after quality control), we simulated samples of size 145. We assumed a neutral single-site mutation rate of $10^{-9}$ per bp per generation [36] and a recombination rate ($\rho$) of $5\times10^{-7}$ centimorgans/bp (cM/bp) [37]. We excluded regions with $\rho < 5x10^{-7}$ cM/bp in the DGRP data to avoid examining regions with elevated LD due to low recombination, as this is at the low end of recombination rates observed in Drosophila [37].

While the decay in pair-wise LD appears to match the data for very short distances for most demographic models, LD in DGRP data is substantially elevated at longer distances, an observation consistent with Przeworski *et al*. [38]. The elevation in LD observed in the data is indicative of either linked positive selection driving haplotypes to high frequency, a lack of fit of current demographic models to the data, or both. Simulations under the most realistic demographic model, admixture [32], have the fastest decay in LD reflecting the increase in the number of haplotypes in an admixture sample as compared to any other model tested (Figure S2). This is likely because admixture models that are fit to diversity statistics in the data generate more haplotypes in the bottleneck population since in this case the same haplotype is unlikely to be sampled independently in both bottlenecks. In contrast LD measured in the constant $N_e$=$10^6$ demographic scenario decays more slowly than in any other demographic scenario because the SNP density per basepair is almost half the density observed under any other model or in the DGRP data (Table S3).

Below we will be defining haplotype analysis windows in terms of the number of SNPs. The lower SNP density of the constant $N_e$=$10^6$ model effectively increases the analysis window size in terms of bps when defining the windows in terms of SNPs, and thus is conservative for the purposes of detecting selection because the recombination rate under this model is artificially increased. For this reason, we chose to use the constant $N_e$=$10^6$ model for the subsequent simulations of neutrality and selective sweeps.

To understand how LD compares in simulations of neutral demography and selective sweeps of varying softness arising from *de novo* mutations and SGV, we visualized sample haplotype frequency spectra in window sizes of 400 SNPs, corresponding to the approximately 10kb over which long-range LD persists in DGRP data (Figure 2). We can estimate the lower bound of selection strength for the sweeps that can be detected in such windows. The footprint of a hard selective sweep extends over approximately $s/[\log(N_e s)\rho]$ basepairs, where $s$ is the selection strength, $N_e$ the population size, and $\rho$ the recombination rate [14,15,39]. Sweeps with a selection coefficient of $s$ = 0.05% or greater are thus likely to affect haplotype homozygosity over 10 kb windows in such areas, assuming a recombination rate of $5\times10^{-7}$ cM/bp, As the recombination rate increases, only selective sweeps with $s$>0.05% should be observed in windows of 10kb. Previous results suggest that selection coefficients of ~1% are common in Drosophila [6,40]. Therefore, our choice of window size will bias us towards detecting stronger selection but should not unduly limit our power to detect adaptation in *D. melanogaster*.

To visualize sample haplotype frequency spectra, we simulated incomplete and complete sweeps of varying softness with frequencies of the adaptive mutation (*PF*) at 0.5 or 1 at the time when selection ceased (Note that below we will investigate a larger number of scenarios, focusing on the effects of varying selection strength and the decay of sweep signatures with time). As can be seen in Figure 3, most haplotypes in neutral demographic scenarios are unique, while selective sweeps generate multiple haplotypes at substantial frequencies. Our plot of the haplotype frequency spectra and the expected numbers of adaptive haplotypes show that for sweeps arising from *de novo* mutations and SGV, sweeps become clearly soft and multiple frequent haplotypes are evident in the sample when $\theta_A \geq 1$ and the starting partial frequency is $\geq 10^{-4}$ (100 alleles in the population), respectively. In both cases, sweeps become monotonically softer as $\theta_A$ and partial frequency of the adaptive allele increases. These results conform to the expectations derived in [17].

The increase of haplotype population frequencies in both hard and soft sweeps can be captured using haplotype homozygosity [25,27]. If $p_i$ is the frequency of the $i^{th}$ most common haplotype in a sample, and $n$ is the number of observed haplotypes, then haplotype homozygosity is defined as $H1 = \Sigma_{i=1,\ldots n}\, p_i^2$. We can expect H1 to be particularly high for hard sweeps, with only one adaptive haplotype at high frequency in the sample Figure 4A. Thus, H1 is an intuitive candidate for a test of neutrality versus hard sweeps, where the test rejects neutrality for high values of H1. A test based on H1 may also have acceptable power to detect soft sweeps in which only a few haplotypes in the population are present at high frequency. However, as sweeps become softer and the number of sweeping haplotypes increases, the relative contribution of individual haplotypes towards the overall H1 value decreases, and the power of a test based on H1 is expected to decrease.

To have a better ability to detect hard and soft sweeps using homozygosity statistics, we developed a modified homozygosity statistic, $H12 = (p_1 + p_2)^2 + \Sigma_{i>2}\, p_i^2 = H1 + 2p_1p_2$, in which the frequencies of the first and the second most common haplotypes are combined into a single frequency (Figure 4B). A statistical test based on H12 is expected to be more powerful in detecting soft sweeps than H1 because it combines frequencies of two similarly abundant haplotypes into a single frequency, while for hard sweeps, the combination of the frequencies of the first and second most abundant haplotypes should not change homozygosity substantially. We also considered a third test statistic, H123, which combines frequencies of the three most prevalent haplotypes in a sample into a single haplotype and then computes homozygosity.

**Ability of H12 to detect selective sweeps**

To assess the ability of H12 to detect sweeps of varying softness and to distinguish positive selection from neutrality, we measured H12 in simulated sweeps arising from both *de novo* mutations and SGV while varying $s$, $PF$, and the time since the end of the sweep, $T_E$, measured in units of $4N_e$ generations in order to model the decay of a sweep through recombination and mutation events over time. We first investigate the behavior of H12 under different selective regimes and then investigate its power in comparison with the popular haplotype statistic *iHS*.

Figure 5A shows that for complete and incomplete sweeps with $s = 0.01$ and $T_E = 0$, H12 monotonically decreases as a function of $\theta_A$ over the interval from $10^{-2}$ to $10^2$. When $\theta_A \leq 0.5$, many sweeps are hard and H12 values are high. When $\theta_A \approx 1$, and practically all sweeps are soft, but not yet extremely soft, H12 retains much of its power. However, for $\theta_A > 10$, where sweeps are extremely soft, H12 decreases substantially. Similarly, H12 is maximized when the starting frequency of the allele is $10^{-6}$ (one copy of the allele in the population generating hard sweeps from SGV) and becomes very small as the frequency of the adaptive allele increases beyond $>10^{-3}$ (>1000 copies of the allele in the population) (Figure 5B). Therefore, H12 has reasonable power to detect soft sweeps in samples of hundreds of haplotypes, as long as they are not extremely soft, but remains somewhat biased in favor of detecting hard sweeps.

H12 also increases as the partial frequency of the sweep ($PF$) increases from 0.5 to 1 (Figures 3A and B) and as the selection strength increases from 0.001 to 0.1 (Figures 5C and D). We observe a 'hardening' of sweeps with smaller values of $s$ (i.e. 0.001) because fewer adaptive alleles reach establishment frequency with a weaker selection coefficient in comparison with sweeps with higher selection coefficients, and so sweeps with a high $\theta_A$ or starting partial frequency of the adaptive allele have high H12 values. Figures 5E and F further show that incomplete and complete sweeps decay with time due to recombination and mutation events, resulting in monotonically decreasing values of H12 with time. As expected, this suggests that H12 has the most power to detect recent sweeps driven by strong selection.

We also assessed the ability of H12 in detecting selective sweeps as compared to H1 and H123 by calculating the values of H1, H12, and H123 for sweeps generated under the parameters $s = 0.01$, $T_E = 0$ and $PF$=0.5. H12 consistently, albeit modestly, increases the homozygosity for younger sweeps as compared to H1 (Figure S3). In comparison to H12, H123 increases the homozygosity very marginally and we chose not to use this statistic in our study.

Finally, we compared the abilities of H12 and *iHS* (integrated haplotype score), a haplotype-based statistic designed to detect incomplete hard sweeps [25,26], to detect hard and soft sweeps. We created receiving operator characteristic (ROC) curves which plot the true positive rate (TPR) of correctly rejecting neutrality in favor of a sweep (hard or soft) given that a sweep has occurred versus the false positive rate (FPR) of inferring a selective sweep, when in fact a sweep has not occurred.

In our simulations of selective sweeps we used $\theta_A = 0.01$ as a proxy for scenarios generating almost exclusively hard sweeps, and $\theta_A = 10$ as a proxy for scenarios generating almost exclusively soft sweeps. We chose $\theta_A = 10$ for soft sweeps because this is the highest $\theta_A$ value with which H12 can still detect sweeps before substantially losing power given our window size of 400 SNPs and sample size of 145. We modeled incomplete sweeps with $PF = 0.1$, 0.5, and 0.9, with varying times since selection had ceased of $T_E = 0$, 0.001, and 0.01 in units of $4N_e$ generations. We simulated sweeps under three selection coefficients, $s = 0.001$, 0.01, and 0.1.

Figures 6 and S4 show that our test based on H12 and the test based on *iHS* have similar power for the detection of hard sweeps, although in the case of old sweeps when $T_E = 0.01$, *iHS* performs slightly better than H12 for the detection of hard sweeps when $s \geq 0.01$. Overall, H12 substantially outperforms *iHS* in detecting soft sweeps when selection is sufficiently strong and the sweeps are sufficiently young. As sweeps become old, neither statistic can detect them well, as expected.

**Haplotype homozygosity scans of DGRP data**

We applied H12 to DGRP data to identify regions of the *D. melanogaster* genome undergoing positive selection. In our application, we controlled for several sources of artifacts including unusually high H12 values arising under various neutral demographic scenarios, strains that share identity by descent, and genomic inversions among the strains. We also reran our scan in additional data sets of the same North Carolina population to verify our results. We used an outlier approach to identify the 50 most empirically extreme H12 peaks in our data. Among our top 3 candidates were the three previously confirmed positively selected loci at *Ace*, *CYP6g1*, and *CHKov1*. We also ran the *iHS* statistic on DGRP data, which should have similar power as H12 in detecting some selective events, to determine the concordance between the two approaches and validate our findings.

**H12 scan of DGRP data**

We applied the H12 statistic to DGRP data in sliding analysis windows of 400 SNPs with the centers of each window iterated by 50 SNPs. In each analysis window, we constructed a haplotype frequency spectrum by grouping together identical haplotypes (Methods). Based on this spectrum, we calculated H12 in each window.

To assess whether the observed H12 values in the DGRP data along the four autosomal arms are unusually high as compared to neutral expectations, we estimated the expected distribution of H12 values under each of the six neutral demographic models. Figure 7 shows that genome-wide H12 values in DGRP data are substantially elevated as compared to expectations under any of the six neutral demographic models. In addition, there is a long tail of outlier H12 values in the DGRP data suggestive of recent strong selective sweeps.

To identify regions of the genome with H12 values significantly higher than expected under neutrality, we calculated critical values ($H12_o$) under each of the six neutral models based on a 1-per-genome false discovery rate (FDR) criterion. Our test rejects neutrality in favor of a selective sweep when $H12 > H12_o$ (Methods and Supplement). The critical $H12_o$ values under all neutral demographic models are similar to the median H12 value observed in the DGRP data (Table 1), consistent with the observations of elevated genome-wide haplotype homozygosity and much slower decay in long range LD in the DGRP data compared to all neutral expectations.. We focused on the constant $N_e = 10^6$ model because it yields a relatively conservative $H12_o$ value and preserves the most long-range pair-wise LD in simulations.

In order to call individual sweeps, we first identified all windows with $H12 > H12_o$ in the DGRP data set under the constant $N_e = 10^6$ model. We then grouped together consecutive windows as belonging to the same 'peak' if the H12 values in all of the grouped windows were above $H12_o$ for a given model and recombination rate (Methods). We then chose the window with the highest H12 value among all windows in a peak and use this H12 value to represent the entire peak.

We focused on the top 50 peaks with empirically most extreme H12 values, hypothesized to correspond to the strongest and/or most recent selective events (Figure 8A). The windows with the highest H12 values for each of the top 50 peaks are highlighted in Figure 8A. The highest H12 values for the top 50 peaks are in the tail of the distribution of H12 values in the DGRP data (Figure 7) and thus are outliers both compared to all neutral expectations and the empirical genomic distribution of H12 values. We observed peaks that have H12 values higher than $H12_o$ on all chromosomes, but found that there are significantly fewer peaks on chromosome 3L (2 peaks) than the approximately 13 out of 50 top peaks expected assuming a uniform distribution of the top 50 peaks genome-wide ($p = 0.00016$, two-sided binomial test, Bonferroni corrected).

The three peaks with the highest observed H12 values correspond to three well-known cases of positive selection in *D. melanogaster* at the genes *Ace*, *CHKov1*, and *CYP6g1* that were described in the Introduction, confirming that our scan is capable of identifying previously known cases of adaptation. In Table S4, we list all genes that overlap with any of the top 50 peaks. Figures 9A and S5 show the haplotype frequency spectra observed at the top 50 peaks. In comparison, Figure 9B shows the frequency spectra observed under the six demographic models with corresponding critical $H12_o$ values.

We performed several tests to ensure the robustness of the H12 peaks to inversions. We found no significant association between the locations of peaks and locations of inversions except in one case (Supplementary Text, Table S5A). Similarly, our test for LD between strains comprising haplotype groups in each peak and strains carrying inversions did not result in any significant association (Table S5B). Figure S6 shows that the there continues to be an elevation and long tail of H12 values in DGRP data even after the removal of regions overlapping major cosmopolitan inversions.

We reran the scan in three different data sets of the same population to ensure that no unaccounted population substructure and variability in sequencing quality confounded our results (Supplementary Text). In all scans, we recovered several of our top peaks. First, we reran the scan in the Drosophila Population Genomics Project (DPGP) data set [41], consisting of 40 strains from the original DGRP data set (Figure S7A). Despite the much smaller sample size of the DPGP data set, we recovered 13 of the highest peaks in the repeated scan, 10 of which are among the top 15 in the DGRP scan. We also reran the scan in version 2 of the DGRP data set [31], which became available during the course of our analysis (Figure S7B). In this new version, we down sampled from 205 to 145 strains to match the sample depth of our original dataset. Since the DGRP data set has been found to have at least five pairs of strains with high genome-wide identity by descent (IBD) values [42] which could contribute to high homozygosity values, we excluded at least one strain from each such pair with high IBD values in this second scan. This scan recovered 34 of our top 50 peaks. We also reran the scan in the 63 strains that were part of the DGRP v2 data set but were not part of the original DGRP data set and recovered 11 peaks observed in the DGRP data (Figure S7C), despite a much smaller sample size. Finally, we sub-sampled the DGRP data set to 40 strains 10 times and plotted the resulting distributions of H12 values (Figure S8). In comparison to H12 distributions observed in the six tested neutral demographic models also sampled at 40 strains, there continues to be an elevation of genome-wide H12 values and a long tail, indicating that the signals observed in the 145-strain DGRP data set are population-wide and are not driven by a sub population.

We scanned chromosome 3R using H1 and H123 as our test statistics in order to determine the impact of our choice of grouping the two most frequent haplotypes together in our H12 test statistic on the location of the identified peaks (Figure S9). We find that the locations of the identified peaks are similar with all three statistics, but that some smaller peaks that cannot be easily identified with H1 are clearly identified with H12 and H123, as expected.

**_iHS_ scan of DGRP data**

We applied the *iHS* statistic as described in Voight et al. 2006 to all SNPs in the DGRP data to determine the concordance in candidates identified by iHS and H12 (Methods). Briefly, we searched for 100 kb windows that have an unusually large number of SNPs with standardized *iHS* values (|*iHS*|) > 2. The positive controls *Ace*, *Cyp6g1*, and *CHKov1* are located within the 95 top 10% iHS 100 kb windows (Figure 8B), validating this approach.

We further overlapped the top 50 H12 peaks with the 95 top 10% iHS 100Kb windows to determine how often a candidate region identified in the H12 scan is identified in the *iHS* scan and vice versa. We defined an overlap as the non-empty intersection of the two genomic regions defining the boundaries of a peak in the H12 scan and the non-overlapping 100Kb windows used to calculate enrichment of |*iHS*| values. We found that 18 H12 peaks overlap 28 |*iHS*| 100Kb enrichment windows. The overlap confirms that many of the peaks identified in both scans are potentially true selective events but that the two approaches are not entirely redundant.

**Distinguishing hard and soft sweeps based on the statistic H2/H1**

Our analysis of H12 haplotype homozygosity and decay in long range LD in DGRP data suggests that extreme outliers in the H12 DGRP scan are in locations of the genome that may have experienced recent and strong selective sweeps. The visual inspection of the haplotype spectra of the top 10 peaks in Figure 9A and the remaining 40 peaks in Figure S5 reveals that peaks contain many haplotypes at substantial frequency. These spectra do not appear similar to those generated by hard sweeps in Figure 3 or extreme outliers under neutrality in Figure 9B but instead visually resemble incomplete soft sweeps with *s*=0.01 and *PF*=0.5 either from *de novo* mutations with $\theta_A$ between 1 and 20 or from SGV starting at partial frequencies of $5x10^{-5}$ to $5x10^{-4}$ (Figure 3). The sweeps also appear to become softer as H12 decreases, consistent with our expectation that H12 should lose power for softer sweeps.

In order to gain intuition about whether the haplotype spectra of the top 50 peaks can be easily generated by hard sweeps versus soft sweeps under several evolutionary scenarios, we developed a new homozygosity statistic, H2/H1, where $H2 = \Sigma_{i>1} p_i^2 = H1 - p_1^2$ is haplotype homozygosity calculated using all but the most frequent haplotype (Figure 4C). We expect H2 to be lower for hard sweeps than for soft sweeps because in a hard sweep, only one adaptive haplotype is expected to be at high frequency [43] and the exclusion of the most common haplotype should reduce haplotype homozygosity precipitously. When the sweep is soft, however, multiple haplotypes exist at high frequency in the population and the exclusion of the most frequent haplotype should not decrease the haplotype homozygosity to the same extent. Conversely H1, the homozygosity calculated using all haplotypes, is expected to be higher for a hard sweep than for a soft sweep as we described above. The ratio H2/H1 between the two should thus increase monotonically as a sweep becomes softer, thereby offering a summary statistic that in combination with H12 can be used to test whether the observed haplotype patterns are likely to be generated by hard or soft sweeps. Note that we intend H2/H1 to be measured near the center of the sweep where H12 is the highest, otherwise further away from the sweep center mutations and recombination events will decay the haplotype signature and hard and soft sweep signatures may look indistinguishable.

**Comparison of the top 50 peaks with sweeps of varying softness using H12 and H2/H1 as summary statistics**

To assess the behavior of H2/H1 as a function of the softness of a sweep, we measured H2/H1 in simulated sweeps of varying softness arising from *de novo* mutations and SGV with various $s$, $PF$, and $T_E$ values. Figure 10 shows that H2/H1 has low values for sweeps with $\theta_A \leq 0.5$ or when the starting partial frequency of the sweep is $<10^{-5}$, when sweeps are mainly hard. As a sweep becomes softer, H2/H1 values approach one because no single haplotype dominates the haplotype spectrum. In the case of sweeps arising from *de novo* mutations, H2/H1 values do not depend on the ending partial frequency of the sweep (*PF)* or selection strength of the sweep. However, in the case of sweeps arising from SGV, sweeps with higher selection strengths have higher H2/H1 values reflecting the hardening of sweeps for smaller $s$ values. Both sweeps from *de novo* mutations and SGV have higher H2/H1 values for older sweeps reflecting the decay of the haplotype frequency spectrum over time.

While hard sweeps and neutrality cannot easily generate both high H12 and H2/H1 values, soft sweeps can do both. In Figure 11 we assess the range of H12 and H2/H1 values expected under hard and soft sweeps. To compare the likelihood of a hard versus soft sweep generating a particular pair of H12 and H2/H1 values, we calculated Bayes factors: BF = $P(H12_{obs}, H2_{obs}/H1_{obs} | \text{Soft Sweep})/P(H12_{obs}, H2_{obs}/H1_{obs} | \text{Hard Sweep})$. We approximated BFs using an approximate Bayesian computation (ABC) approach under which the nuisance parameters selection coefficient ($s$), partial frequency ($PF$), and age ($T_E$) are integrated out by drawing them from uniform prior distributions: $s \sim U[0,1]$, $PF \sim U[0,1]$, and $T_E \sim U[0,0.001] \times 4N_e$. We stated the hard and soft sweep scenarios as point hypotheses in terms of the $\theta_A$ value generating the data. Specifically, we assumed that hard sweeps are generated under $\theta_A = 0.01$. For soft sweeps, we generated sweeps of varying softness by using $\theta_A$ values of 5, 10, and 50. Note hard and soft sweeps can also be simulated from SGV with various starting frequencies of the beneficial allele, but for the purposes of generating hard sweeps with a single sweeping haplotype versus soft sweeps with multiple sweeping haplotypes, simulations from either SGV or *de novo* mutations are equivalent.

The panels in Figure 11 show BFs calculated under several evolutionary scenarios for a grid of H12 and H2/H1 values. All panels in Figure 11 show that hard sweeps are common when H2/H1 values are low for most H12 values tested. For very low H12 (<0.05) values, when sweeps display low haplotype homozygosity to begin with and are difficult to detect with H12, both hard and soft sweeps are likely for a wide range of H2/H1 values. Soft sweeps are common for any high H2/H1 values conditional on H12 being sufficiently high when simulating soft sweeps with $\theta_A = 10$ and 5 (Figures 11A and B). However, soft sweeps generated with $\theta_A = 50$ are too soft to produce high H12 values, confirming our results in Figure 5, and as a consequence hard sweeps are common for high H12 values regardless of H2/H1 values under this scenario (Figure 11C). In Figures 11A, D and E, the recombination rate is varied, and a comparison of these panels show that the recombination rate has little impact on the space where hard sweeps can be expected to be more likely. Figure 11F shows that simulations under admixture increase support for soft sweeps in regions of the space already in support of soft sweeps generated under the constant $N_e$=$10^6$ demographic scenario (Figures 11A-E). Figure 10 shows that there is clearly a dependency between H12 and H2/H1 and that both values need to be taken into account when determining the softness of a peak. In particular, H2/H1 is most informative when applied to regions of the genome with high H12 values.

Overlaid on all panels in Figure 11 are the H12 and H2/H1 values at the top 50 peaks. Note that in almost all the cases, all top 50 peaks have H12 and H2/H1 values that are easiest explained by soft sweeps. In order to more explicitly test each candidate sweep for its compatibility with a hard and soft sweep model, we generated hard sweeps with $\theta_A = 0.01$ and soft sweeps with a maximum a posteriori $\theta_A$ value ($\theta_A^{MAP}$), i.e. our best estimate of the softness for a particular peak. We used an ABC method to infer the $\theta_A^{MAP}$ for each peak by sampling the posterior distribution of $\theta_A$ conditional on the observed values $H12_{obs}$ and $H2_{obs}$ /$H1_{obs}$ from a candidate sweep (Supplement). All $\theta_A^{MAP}$ values inferred for the top 50 peaks were significantly greater than 1 with the smallest being 6.8, suggesting that soft sweeps would be commonly generated under any of the $\theta_A^{MAP}$ values estimated (Figure 3). We used recombination rates estimated for each peak [37] and simulated the data under the constant population size model with $N_e = 10^6$ for computational feasibility. Among our top 50 peaks, we found strong evidence in support of soft sweeps in all 50 cases (BF > 10), very strong evidence in 47 cases (BF > 30), and almost decisive evidence (BF > 98) in 44 cases (Table S3). Taken together, these results provide evidence that soft sweeps most easily explain the signatures of multiple haplotypes at high frequency observed at the top 50 peaks.

## DISCUSSION

In this study, we investigated the genome-wide landscape of selective sweeps in a North American population of *D. melanogaster*. In contrast to previous studies, we employed two newly developed haplotype statistics that have substantial power to detect both hard and soft sweeps and to differentiate them from each other. We find compelling evidence of a substantial number of recent and strong selective sweeps in the North Carolina population of *D. melanogaster* and further find that practically all such sweeps appear to display signatures of soft rather than hard sweeps.

To detect recent and strong adaptation, we used H12, which measures haplotype homozygosity in an analysis window after combining the frequencies of the two most abundant haplotypes into a single frequency. Unlike *iHS*, another commonly used haplotype statistic, H12 is capable of detecting both hard and soft sweeps with similar power, as long as the sweeps are not too soft. If sweeps arise from *de novo* mutations, the upper bound of $\theta_A$ we can reliably detect is ~10, and if sweeps arise from SGV, the advantageous variant must be at low frequency ($< 10^{-4}$ in a population of $10^6$) at the onset of selection (Figure 5).

We scanned the *D. melanogaster* genome with H12 over windows of 400 SNPs (~10kb) in regions of recombination ($\rho$) greater than $5 \times 10^{-7}$ cM/bp (86.2% of the genome). Application of H12 with a window of this size gives us power to detect primarily recent and strong adaptation (Figures 5, 6, and S4), since sweeps driven by weak selection do not generate haplotypes long enough to span the whole 400 SNP window. The power of the H12 scan also decays rapidly with the age of the sweep as recombination breaks down common haplotypes. We conducted extensive simulations to show that this choice of window size indeed represents a good trade-off between detecting recent strong selection and having a low rate of false positives under a broad range of demographic scenarios.

We chose to use windows defined by a constant number of SNPs rather than windows of constant physical or genetic length in order to simplify the statistical analysis. This is because windows of constant physical or genetic length tend to have varying SNP density, and thus also varying distributions of haplotypes even under neutrality. Our choice of a fixed number of SNPs avoids this source of noise, but it does bring up the question of whether we end up selecting regions that have particularly low recombination rates or high SNP densities and thus short windows in terms of base pairs or genetic map length. We made sure to avoid the first pitfall by analyzing only the windows with reasonably high recombination rates ($\rho \geq 5x10^{-7}$ cM/bp) and by using conservative thresholds for the significance cutoffs. We also confirmed that the peaks with the highest H12 values do not have particularly high SNP densities per kb (data not shown). We were further concerned that the use of SNP windows would bias us against detecting complete hard sweeps. However, our simulations showed that this was not the case (Figure 5).

In order to control for unexpectedly high H12 values in the DGRP data arising from neutral demographic processes, we generated a distribution of H12 values from simulations under six neutral demographic models. These models include an admixture model proposed by Duchen *et al*. (2013) [32], a variant of the admixture model with a bottleneck from one of the mixing populations [32], a constant $N_e = 10^6$ model, a constant $N_e = 2.7x10^6$ model fit to Watterson's $\theta_W$ inferred from short intronic regions of the genome, and two bottleneck models fit to short intronic regions using the software DaDi [35]. Compared to H12 distributions generated in neutral demographic simulations, we observed an elevation and long tail of high H12 values in the DGRP data (Figure 7). Surprisingly, the 1-per-genome FDR H12 critical values calculated for all demographic models were not significantly higher than the median H12 value observed in the data. In addition, a comparison in the decay of pair wise LD in DGRP data and in simulations under each neutral demographic model revealed that long-range LD in DGRP data is substantially elevated (Figure 2) despite a fit of most demographic models to the data in terms of $S$, $\pi$, and short-range LD (Table S3).

Currently there is no known demographic model that fits the DGRP data in terms of all summary statistics including $S$, $\pi$, haplotype homozygosity, short-range LD and long-range LD. Background selection (BGS), which Comeron [44] has shown to have pervasive effects genome-wide, could impact estimates of S and π and thus should be certainly accounted for in demographic inferences. However, Enard *et al*. [45] showed that haplotype homozygosity is not elevated due to BGS alone. One possible explanation for the genome-wide elevation in haplotype homozygosity is that *D. melanogaster* has undergone a large number of recurrent selective sweeps in its past which have not fully decayed to levels of homozygosity observed under strict neutrality. This is a scenario that needs to be further investigated, and it is clear that further development of the demographic model of North American *D. melanogaster* is required. Importantly, however, the top 50 H12 peaks we discovered in the DGRP data are outliers not only under all demographic models we tested but also outliers relative to the empirical genomic H12 distribution.

Our top three candidates correspond to the well-known cases of soft selective sweeps arising from *de novo* mutations and SGV at the loci *Ace, Cyp6g1,* and *CHKov1* [9,11,13] as described in the Introduction. The recovery of these positive controls validates that our method can identify sweeps arising from both *de novo* mutations and SGV. Note that our method does not have the ability to differentiate whether a soft sweep arose from standing variation or from multiple *de novo* mutations. Peter *et al*. (2012) [46] developed an approximate Bayesian computation method that distinguishes a given sweep as either resulting from a single *de novo* mutation (generating a signatures of a hard sweep) or from standing genetic variation (generating a signature of a soft sweeps). However, their method can only be applied to selective sweeps that have already been identified with pre-existing methods and does not distinguish soft sweeps from multiple *de novo* mutations versus those from standing variation.

In addition to H12, we ran *iHS* on the DGRP data and recovered 18 of the top 50 peaks, including the three positive controls, demonstrating the validity of both methods and that the two methods are not entirely redundant (Figure 8B). We further performed a number of checks to assess the robustness of our top 50 peaks to unaccounted substructure in the data. First, we tested for enrichments for peaks in regions of the genome with inversions because inversions can result in elevated levels of homozygosity due to suppression of recombination especially near breakpoints. Out of the 7 inversions tested, we found that one inversion on chromosome 3R had an enrichment of peaks (Table S5A). However, we also checked whether any of the strains comprising the main haplotype clusters in our top peaks were correlated with inversions on the same chromosome and could not find any such correlation for any of the peaks (Table S5B) (Supplement). Figure S6 shows that even after the removal of major cosmopolitan inversions, there continues to be an elevation and long tail of high H12 values.

The DGRP data set contains several pairs of strains with high genome-wide IBD values suggesting sibling and cousin relationships [42,47]. We repeated the scan excluding one individual of each of these pairs in an updated DGRP v2 data set and found that 40 peaks in the new scan overlapped 34 of the original 50 peaks, most of which were top-ranking (Figure S7B). This result suggests that our method is robust to any contribution of homozygosity of related individuals to the homozygosity observed in the peaks. We also reran the scan in the DPGP data set consisting of 40 of the original 162 lines used in the DGRP data set, and even with such a small sample size we still recovered 13 of our top peaks, most of which were highly ranking as well (Figure S7A).

We further validated our results in 63 strains in the DGRP v2 data set that were non-overlapping with the DPGP data used for our scan and again recovered 11 of our original peaks despite the small sample size and increased amount of missing data in these strains relative to the strains used for the DGRP analysis (Figure S7C). Finally, we sub-sampled the DGRP data set 10 times to 40 strains and compared the resulting distributions of genome-wide H12 values observed in each sample and found that there continued to be an elevation of high H12 values and long tail relative to any of the neutral demographic models tested (Figure S8). Taken together, these scans in independent data sets confirm that our results are robust to hidden substructure in the data.

The visual inspection of the haplotype spectra at the top 50 peaks in Figures 9 and S5 show that there are multiple haplotypes present at high frequency at all the top peaks. The patterns observed at these peaks do not seem consistent with that of a hard sweep, where only one haplotype is at high frequency, or with neutrality, where even in the extreme tails of the H12 distribution under any neutral demographic scenario we are unable to generate high enough H12 to values to match outliers in the data. Rather, the sweeps have signatures reminiscent of incomplete soft sweeps, as depicted in Figure 3, where the combined frequencies of the first and second most frequent haplotypes reach up to 30%.

To assess whether the top 50 peaks can be more easily generated by hard versus soft sweeps, we developed a second statistic, H2/H1, which is a ratio of homozygosity values calculated without and with the most frequent haplotype in a sample. We demonstrate that this statistic has a monotonic increasing relationship with the softness of a sweep (Figure 10). Taken together with H12, both statistics can be informative in determining the softness of a sweep. Specifically, hard sweeps can generate high values of H12 in a window centered on the adaptive site but cannot at the same time generate high H2/H1 values in the window, while soft sweeps can generate both high H12 and H2/H1 values in such a window. Note that in order to differentiate hard and soft sweeps with reasonable power, H2/H1 can only be applied in cases where H12 values are already high and there is a strong evidence of a sweep.

Why should we expect low H2/H1 values for hard sweeps? In a hard sweep, we expect one adaptive haplotype to be at high frequency and the variants of that adaptive haplotype that arose from early mutation or recombination events to be at much lower frequencies. Specifically, we expect the most abundant adaptive haplotype to be approximately $s/\mu$ times more prevalent than its most frequent variant [43], where $s$ is the selection coefficient of the sweep and $\mu$ is the combined mutation and recombination rate over the analysis window. For example, in our analysis, we utilize a window size of 400 segregating sites corresponding to approximately 10kb in the DGRP data. This is a reasonable choice for our analysis because assuming a mutation rate of $2*10^{-9}$/bp/gen and a recombination rate of $5*10^{-7}$ cM/bp, $\mu$ in the region is approximately $7*10^{-5}$/gen $=10^4$ bp* ($2*10^{-9}$/bp/gen + $5*10^{-7}$ cM/bp). Thus, even for sweeps driven by fairly weak positive selection, e.g. $s = 0.001$, the expected ratio of the most frequent and second most-frequent haplotype is about $10^{-3}/7*10^{-5} \approx 15$. Since this ratio is proportional to $s$, it is expected to become even larger for stronger selection. Thus, variants of the main sweeping haplotype resulting from mutation or recombination events during the sweep are not expected to contribute substantially to haplotype homozygosity. Therefore, H1, and H2/H1 values should be low most of the time for hard sweeps as long as both are calculated at or near the center of the peak.

We tested how easily H12 and H2/H1 values for the top 50 peaks can be generated under hard and soft sweeps in a number of evolutionary scenarios. In Figure 11 we calculate BFs on a grid of H12 and H2/H1 values to determine the range of H12 and H2/H1 values most likely to be observed under hard versus soft sweeps. As can be seen in all evolutionary scenarios presented in Figure 11, as long as H12 is sufficiently high, when H2/H1 is low, hard sweeps are common, and when H2/H1 is high, soft sweeps are common. However, when H12 is very low, *i.e.*, when there is little evidence for a sweep to begin with, a wider range of H2/H1 values are compatible with hard sweeps. The dependency of H12 and H2/H1 demonstrate that the two statistics must be applied jointly to infer the softness of a peak, and only in cases when H12 is high enough to be distinguishable from neutrality.

We overlaid the H12 and H2/H1 values observed for the top 50 peaks on the grid of BFs measured for all H12 and H2/H1 values. In all scenarios, the top 50 peaks have H12 and H2/H1 values that lie in regions with high BFs corresponding to soft sweeps, regardless of the values of $\theta_A$, recombination rate, and demographic model used for simulations. Further testing of each peak with the exact recombination rate observed at each peak and the maximum a posteriori $\theta_A$ best fitting each peak reconfirmed our results that the top 50 peaks show signatures most easily generated by soft sweeps commonly and hard sweeps rarely, if at all.

Note that our choice to simulate hard and soft sweeps under the constant $N_e$=$10^6$ demographic model makes our analysis conservative for the purposes of rejecting the hard sweep scenario because the lower SNP density in the $N_e$=$10^6$ model (Table S3) as compared to DGRP data effectively increases the analysis window size in terms of base pairs, and by extension, also increases the number of recombination events each window experiences. Thus, hard sweeps should look "softer" under this choice of demographic model. Even still, soft sweeps and not hard sweeps seem to more easily explain the signatures at our top 50 peaks.

Our results suggest that recent and strong adaptation generated common signatures of soft selective sweeps in the Drosophila genome. Interestingly we do not see any complete sweeps or sweeps with only one haplotype at high frequency. One possibility is that some of our top candidates are under balancing selection, and this might prevent sweeps from reaching completion, or that the sweeping variants are beneficial in some but not all populations or under some but not all environmental conditions.

If soft sweeps are indeed common in *D. melanogaster*, then adaptation must act on SGV at low enough frequencies to generate high enough H12 values to be detected or multiple *de novo* adaptive mutations entering the population simultaneously. A reason why the adaptive mutation rate may be high enough to generate common signatures of soft sweeps is that the population size relevant for recent adaptation could be closer to the census population size at the time of adaptation as compared to the commonly assumed value of $N_e = 10^6$ for the effective population size in *D. melanogaster*. A value of $N_e = 10^6$ for the effective population size in *D. melanogaster* is much smaller than the reciprocal of the mutation rate per bp of $10^{-9}$ and suggests that adaptation from *de novo* mutation at single sites is mutation-limited and should generally lead to signatures of hard sweeps. Instead, a more likely scenario is that the adaptation-relevant $N_e$ is much larger than $10^6$ in *D. melanogaster*, especially when we consider the *D. melanogaster* population as a whole, as has been previously argued [9,29]. Another possibility is that in many cases the adaptive mutation rate is much higher than the single nucleotide mutation rate because the mutational target size may be larger than one basepair.

Yet where are the hard sweeps? One possibility is that hard sweeps do exist but are driven by weak selection, and thus are missed by our scan. Indeed, Wilson *et al*. [30] argued that sweeps driven by weak selection could become hard even when they occur in populations of large size, in cases where they take a long enough time to increase in frequency such that rare, sharp bottlenecks eliminate all but the highest frequency adaptive allele. Another possibility is that we may be observing signatures of multiple local hard sweeps arising within sub-demes of the Drosophila population or in the ancestral European and African populations prior to admixture, but these would be considered soft sweeps given the population as a whole [48].

It is also possible that hard sweeps were common in the past and degraded over time, while recent adaptation from *de novo* or rare variants produced primarily soft sweeps. While it is possible that hard sweeps correspond to the weaker and older selection events that we lack power to identify, it is reassuring that our method is biased toward discovering the strongest and most recent adaptive events in the genome.

The abundance of signatures of soft sweeps in *D. melanogaster* has important implications for the design of methods used to quantify adaptation. Some methods may work equally well whether adaptation leads to signatures of hard or soft sweeps. For instance, estimates of the rate of adaptive fixation derived from McDonald-Kreitman tests [49] are not expected to be affected by the predominant type of sweeps, because these estimates depend on the rate of fixation of adaptive mutations, and not on the haplotype patterns of diversity that these adaptive fixations generate in their wake. Tests based on the prediction that regions of higher functional divergence should harbor less neutral diversity [40,50,51], are generally consistent with recurrent hard and soft sweeps, as both scenarios are expected to increase levels of genetic draft, and thus reduce neutral diversity in regions of frequent and recurrent adaptation. However, methods that quantify adaptation based on a specific functional form of the dependence between the level of functional divergence and neutral diversity may lead to different conclusions under hard and soft sweeps [40]. Finally, methods that rely on the specific signatures of hard sweeps, such as the presence of a single frequent haplotype [25,26], sharp local dip in diversity [15], or specific allele frequency spectra expected during the recovery after the sweep might often fail to identify soft sweeps [23]. Hence, such methods might give us an incomplete picture of adaptation. Moreover, such methods might erroneously conclude that certain genomic regions lacked recent selective sweeps, which can be problematic for demographic studies that rely on neutral polymorphism data unaffected by linked selection.

Our statistical test based on H12 to identify both hard and soft sweeps and our test based on H12 and H2/H1 to distinguish signatures of hard versus soft sweeps can be applied in all species in which genome-scale, phased polymorphisms data are available and can easily be extended to unphased data as well. Our methods require a sufficiently deep population sample for precise measurement of haplotype frequencies, which is essential for determining whether a haplotype is unusually frequent in the sample. For example, in our DGRP scan, the majority of the 50 highest H12 peaks had a combined frequency of the two most common haplotypes below 30%, while only the top three peaks had a combined frequency of approximately 45%. Furthermore, in order to determine whether an observed H12 value is sufficiently high to suggest that a sweep has occurred in the first place, a robust picture of demographic history and reliable estimates of recombination rates are needed.

Our results provide evidence that signatures of soft selective sweeps were abundant in recent evolution of *D. melanogaster*. Soft sweep signatures may be common in many additional organisms which have a high census population size, including plants, marine invertebrates, insects, microorganisms, and even modern humans when considering very recent evolution in the population as a whole. Indeed, the list of known soft sweeps is large, phylogenetically diverse, and is constantly growing [29]. A comprehensive understanding of adaptation therefore must account for the possibility that soft selective sweeps are a frequent and possibly dominant mode of adaptation in nature.

## METHODS

### Simulations of selection and neutrality

Population samples under selection and neutrality were simulated with the coalescent simulator MSMS [52]. We simulated samples of size 145 to resemble the sample depth of the DGRP data and always assumed a neutral mutation rate of $10^{-9}$ bp/gen [36].

MSMS can simulate selective sweeps both from *de novo* mutation and standing genetic variation. For the *de novo* scenarios, we generated selective sweeps of varying softness by specifying the population parameter $\theta_A = 4N_e\mu_A$ at the adaptive site. For the standing genetic variation scenarios, we specified the initial frequency of the adaptive allele in the population at the onset of positive selection. The adaptive site was always placed in the center of the locus. We assumed co-dominance, whereby a homozygous individual bearing two copies of the advantageous allele has twice the fitness advantage of a heterozygote. To simulate incomplete sweeps we specified the ending partial frequency of the sweep. To simulate sweeps of different age we conditioned on the ending time of selection ($T_E$) prior to sampling.

When simulating selection with the admixture demographic model, it was unfortunately not possible in MSMS to condition on $T_E$. In this particular case, we instead conditioned on the start time of selection in the past and the starting partial frequency of a sweep, with selection continued until the time of sampling. In doing so, we assumed a uniform prior distribution of the start time of selection, U[0 to $3.05\times10^{-4}N_e$] generations, with the upper bound specifying the time of the admixture event.

**Performance analysis of haplotype statistics**

We simulated loci of length $10^5$ bp for sweep simulations with $s < 0.1$ and $10^6$ bp for sweep simulations with $s = 0.1$. For neutral simulations, we simulated loci of length $10^5$ bp. We assumed a constant effective population size of $N_e = 10^6$ and a recombination rate of $5\times10^{-7}$ cM/bp, reflecting the cutoff used in the DGRP analysis.

Our statistics H12 and H2/H1 were estimated over windows of size 400 SNPs centered on the adaptive site. Simulated samples that yielded fewer than 400 SNPs were discarded. For the comparison with *iHS*, we calculated *iHS* values for the SNP immediately to the right of the selected allele, and determined the size of the region by cut-off points at which *iHS* levels decayed to values observed under neutrality. In some simulation runs under the extreme scenario with $s = 0.1$ and $T_E = 0$, *iHS* had not yet decayed to neutral levels at the edges of the simulated sweep. However, this should have only minor impact on the ROC curves.

**Quality filtering of the DGRP data**

The DGRP data set generated by Mackay *et al*. (2012) [31] consists of the fully sequenced genomes of 192 inbred *D. melanogaster* lines collected from a Raleigh, North Carolina population. Reference genomes are available only for 162 lines. Of these 162 lines, we filtered out a further 10% of the lines with the highest number of heterozygous sites in their genomes, possibly reflecting incomplete inbreeding. The IDs of these strains are: 49, 85, 101, 109, 136, 153, 237, 309, 317, 325, 338, 352, 377, 386, 426, 563, and 802. Our final data set consisted of 145 strains.

**Genomic scan for selective sweeps in DGRP using H12**

We scanned the genome using sliding windows of 400 SNPs with intervals of 50 SNPs between window centers and calculated H12 in each window. If two haplotypes differed only at sites with missing data, we clustered these haplotypes together. If multiple haplotypes matched a haplotype with missing data, we clustered the haplotype with missing data at random with equal probability with one of the other matching haplotypes. We treated the heterozygous sites in the data as sites with missing data ("N").

To identify regions with unexpectedly high values of H12 under neutrality, we calculated the expected distribution of H12 values under the admixture, admixture and bottleneck, constant $N_e = 10^6$, constant $N_e = 2.7\text{x}10^6$, severe short bottleneck, and shallow long bottleneck demographic scenarios specified in Figure 1. For each scenario, we simulated ten times the number of independent analysis windows (approximately $1.3\text{x}10^5$ simulations) observed on chromosomes 2L, 2R, 3L, and 3R using three different recombination rates: $10^{-7}$ cM/bp, $5\times10^{-7}$ cM/bp, and $10^{-6}$ cM/bp. All simulations were conducted with locus lengths of $10^5$ basepairs. We assigned a 1-per-genome FDR level to be the 10th highest H12 value in each scenario.

Consecutive windows with H12 values that are above the 1-per-genome-FDR level were assigned to the same peak by the following algorithm: First, we identified the highest H12 value along a chromosome that lies above the 1-per-genome-FDR with a recombination rate greater than $5\times10^{-7}$ cM/bp. We then grouped together all consecutive windows that also lie above the cutoff and assigned these all to the same peak. After identifying a peak, we chose the highest H12 value among all windows in a peak to represent the H12 value of the entire peak. We repeated this procedure for the remaining windows until all analysis windows were accounted for.

**Genomic scan of DGRP data with *iHS***

We scanned the DGRP data using a custom implementation of the *iHS* statistic written by Sandeep Venkataram and Yuan Zhu. *iHS* was calculated for every SNP with a minor allele frequency (MAF) of at least 0.05 without polarization. Any strain with missing data in the region of extended haplotype homozygosity for a particular SNP was discarded in the computation of *iHS*. All *iHS* values were normalized by a distribution of *iHS* values calculated at all SNPs sharing the same MAF as the SNP being normalized. As described in Voight *et al*. [26], we calculated the enrichment of SNPs with standardized *iHS* values > 2 in non-overlapping 100 Kb windows.

**Demographic inference with DaDi**

We fit six simple bottleneck models to DGRP data using a diffusion approximation approach as implemented by the program DaDi [35]. DaDi calculates a log-likelihood of the fit of a model based on an observed site frequency spectrum (SFS).

We estimated the SFS for presumably neutral SNPs in the DGRP using segregating sites in short intron [53]. Specifically, we used every site in a short intron of length less than 86 bps, with 16 bps removed from the intron start and 6 bps removed from the intron end. We projected the SFS for our data set down to 130 chromosomes (after excluding the top 10% of strains with missing data), resulting in 42,679 SNPs out of a total of 738,024 bps.

We specified a constant population size model as well as six bottleneck models with the sizes of the bottleneck ranging from 0.2% to 40% of the ancestral population size. Using DaDi, we inferred three free parameters: the bottleneck time ($T_B$), final population size ($N_F$) and the final population time ($T_F$) (Figure S1 and Table S2). All six bottleneck models produced approximately the same log likelihood values and estimates of $N_F$ and $T_F$. Further, the estimates of $S$ and $\pi$ obtained from simulated data matched the estimates obtained from the observed short intron data (Table S3). Note that the estimate of $T_B$ is proportional to $N_B$, reflecting the difficulty in distinguishing short and deep bottlenecks from long and shallow bottlenecks. We inferred $N_e$ = 2,657,111 ($\approx 2.7 \times 10^6$) for the constant population size model, assuming a mutation rate of $10^{-9}$/bp/generation.

**ABC inference of $\theta_A^{MAP}$ for top 50 peaks**

To infer $\theta_A^{MAP}$ values for the top 50 peaks (Supplement), we assumed uniform distributions for all model parameters in our ABC procedure: The adaptive mutation rate ($\theta_A$) took values on [0,100], the selection coefficient $s$ on [0,1], the partial frequency ($PF$) on [0,1], and the age of the sweep ($T_E$) on [0,0.001]×4$N_e$. We assigned a recombination rate to each peak according to the estimates from Comeron *et al*. (2012) [37] for the specific locus. For the ABC procedure, we binned recombination rates into 5 equally spaced bins. Then, for each peak, we simulated the recombination rate from a uniform distribution over the particular bin its recombination rate fell in. The recombination rate intervals defining the 5 bins were: [$5.42*10^{-7}$, $1.61*10^{-6}$), [$1.61*10^{-6}$, $2.68*10^{-6}$), [$2.68*10^{-6}$, $3.74*10^{-6}$), [$3.74*10^{-6}$, $4.81*10^{-6}$), [$4.81*10^{-6}$, $5.88*10^{-6}$) in units of cM/bp. We assumed a demographic model with constant $N_e = 10^6$ and a non-adaptive mutation rate of $10^{-9}$ bp/gen.

For each peak, we sampled an approximate posterior distribution of $\theta_A$ by finding 1000 parameter values that generated sweeps with H12 and H2/H1 values within 10% of the observed values $H12_{obs}$ and $H2_{obs}/H1_{obs}$ for the particular peak. We calculated the lower and upper 95% credible interval bounds for $\theta_A$ using the 2.5$^{th}$ and 97.5$^{th}$ percentiles of the posterior sample. On each posterior sample, we applied a Gaussian smoothing kernel density estimation and obtained the maximum a posteriori estimate $\theta_A^{MAP}$ for each peak.

We used the same procedure for obtaining approximate posterior distributions of $\theta_A$ and $\theta_A^{MAP}$ estimates under the admixture model. In this case, we used a uniform prior distribution $T_S \sim U[0, 3.05\times10^{-4}]\times N_e$, where $3.05\times10^{-4}N_e$ generations is the time of the admixture event. The prior distributions for parameters other than $T_S$ were the same as for the constant $N_e = 10^6$ model.

**Test of hard versus soft sweeps for the top 50 peaks.**

We used an ABC approach to calculate Bayes factors for a range of H12 and H2/H1 values. We simulated hard sweeps with $\theta_A = 0.01$ and soft sweeps with $\theta_A = 5, 10, 50$, or the $\theta_A^{MAP}$ inferred for a particular peak, depending on the scenario being tested. In the constant $N_e = 10^6$ models shown in Figures 11A–E, selection coefficients, partial frequencies and sweep ages were drawn from uniform distributions: $s \sim U[0,1]$, $T_E \sim U[0, 10^4]\times4N_e$, $PF \sim U[0,1]$. For the admixture model in Figure 11F, the age of the onset of selection from a uniform distribution $T_S \sim U[0, 3.05\times10^{-4}]N_e$ generations, where $3.05\times10^{-4}N_e$ generations corresponds to the time of the admixture event.

We calculated our Bayes factors by taking the ratio of the number of data sets simulated with H12 and H2/H1 values with a Euclidean distance <0.1 from the observed values $H12_{obs}$ and $H2_{obs}/H1_{obs}$ for each set of $10^6$ simulated data sets under soft versus hard sweeps ($10^5$ data sets were generated for explicitly testing each peak with $\theta_A^{MAP}$). We calculated Euclidean distance as follows: $d_i = \sqrt{\left(H12_{obs} - H12_i\right)^2 / Var(H12) + \left(H2_{obs}/H1_{obs} - H2_i/H1_i\right)^2 / Var(H2/H1)}$, where Var(H12) and Var(H2/H1) are the estimated variances of the statistics H12 and H2/H1 calculated using all simulated data sets.

## ACKNOWLEDGMENTS

We thank Noah Rosenberg, Carlos Bustamante, Gavin Sherlock, Hua Tang, Joachim Hermisson, Nick Barton, Kevin Thornton, members of the Petrov Lab, and participants of the Society for Molecular Biology and Evolution conference 2013 for helpful discussions and comments. We thank Pleuni Pennings, Ben Wilson, David Enard, David Lawrie, Arbel Harpak, and Heather Machado for comments on the manuscript. We thank Sandeep Venkataram and Yuan Zhu for their code to measure *iHS* values. We thank Sonali Aggarwal, Greg Ewing, Pablo Duchen, and Aleksandr Arhipov for their additional help with the computational analyses. This work was supported by NIH grants R01 GM100366, R01 GM097415, R01 GM089926 to DAP, and R01 GM081441 to EOB, the NSF Graduate Research Fellowship to NRG, and the HFSP fellowship to PWM. We thank Stanford BioX, Scalegen.com, University of Idaho iBEST (National Center for Research Resources (5P20RR016448-10) and the National Institute of General Medical Sciences (8 P20 GM103397-10) from the National Institutes of Health), University of Texas Ranger, Stanford Center for Genomics and Personalized Medicine, and Stanford Barley computer clusters for providing access to computational resources.

**Figure 1: Neutral demographic models considered.** We estimated LD decay and distributions of haplotype homozygosity in six neutral demographic models for North American *D. melanogaster*. The models considered are as follows: (A) An admixture model as proposed by Duchen *et al*. [32]. (B) An admixture model with the European population undergoing a bottleneck. This was also tested by Duchen *et al*. [32] but the authors found it to have a poor fit. See Table S1 for parameter estimates and symbol explanations for models A and B. (C) A constant $N_e = 10^6$ model. (D) A constant $N_e = 2.7\mathrm{x}10^6$ model fit to Watterson's $\theta_{\mathrm{W}}$ measured in short intron autosomal polymorphism data from the DGRP data set. (E) A severe short bottleneck model and (F) a shallow long bottleneck model fit to short intron regions in the DGRP data set using the software DaDi. See Table S2 for parameter estimates for models E and F. All models except the constant $N_e = 10^6$ model fit the DGRP short intron data in terms of $S$ and $\pi$ (Table S3).

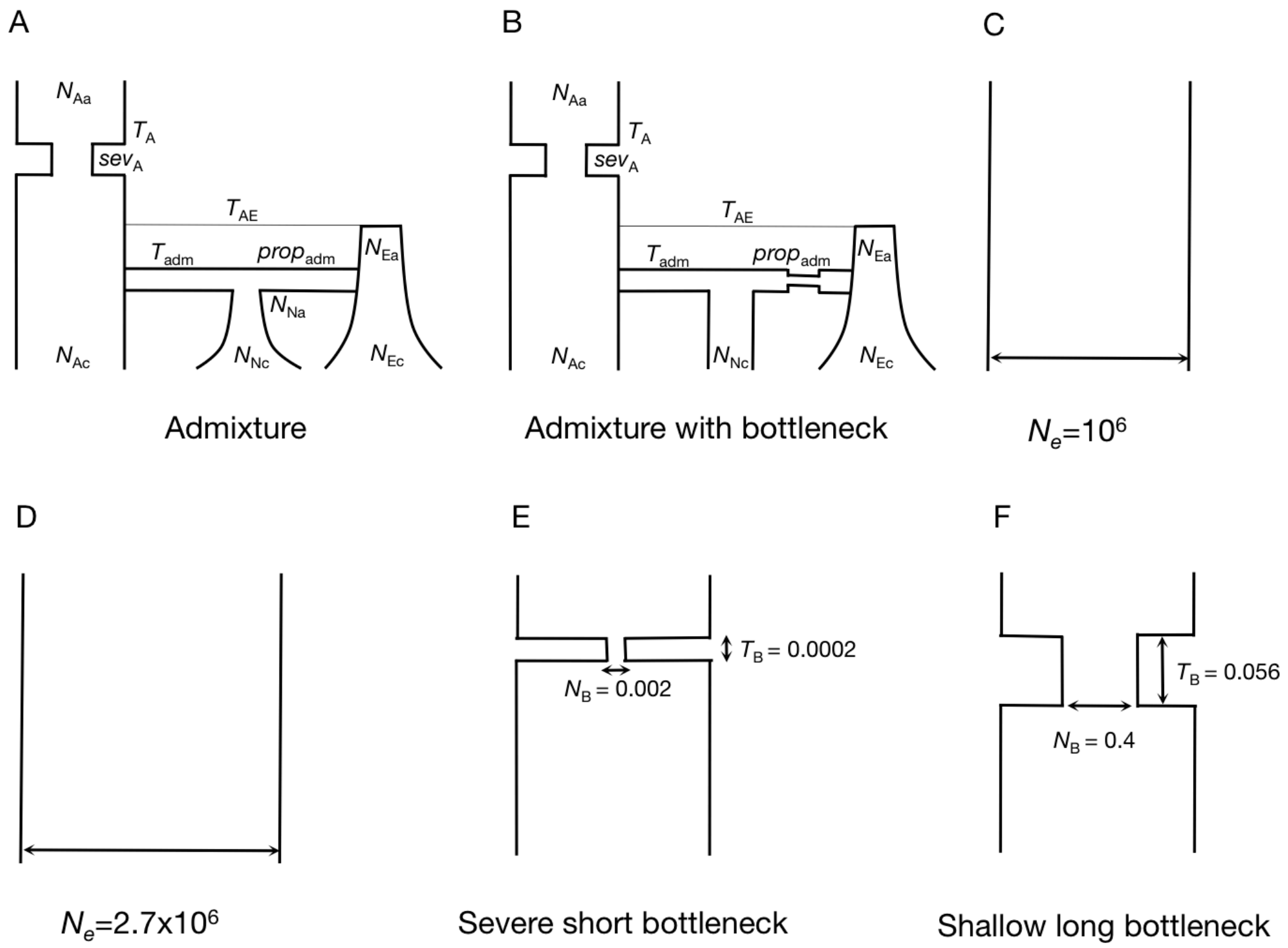

**Figure 2: Elevated long-range LD in DGRP data.** Long range LD in DGRP data is elevated as compared to any neutral demographic model. However, short range LD under most models fit the data well, reflecting the fact that short fragments were used to infer the demographic models in the first place. Pairwise LD was calculated in DGRP data for regions of the *D. melanogaster* genome with $\varrho \geq 5\times10^{-7}$ cM/bp. Neutral demographic simulations were generated with $\varrho = 5\times10^{-7}$ cM/bp. Pairwise LD was averaged over 3x$10^{4}$ simulations in each neutral demographic scenario.

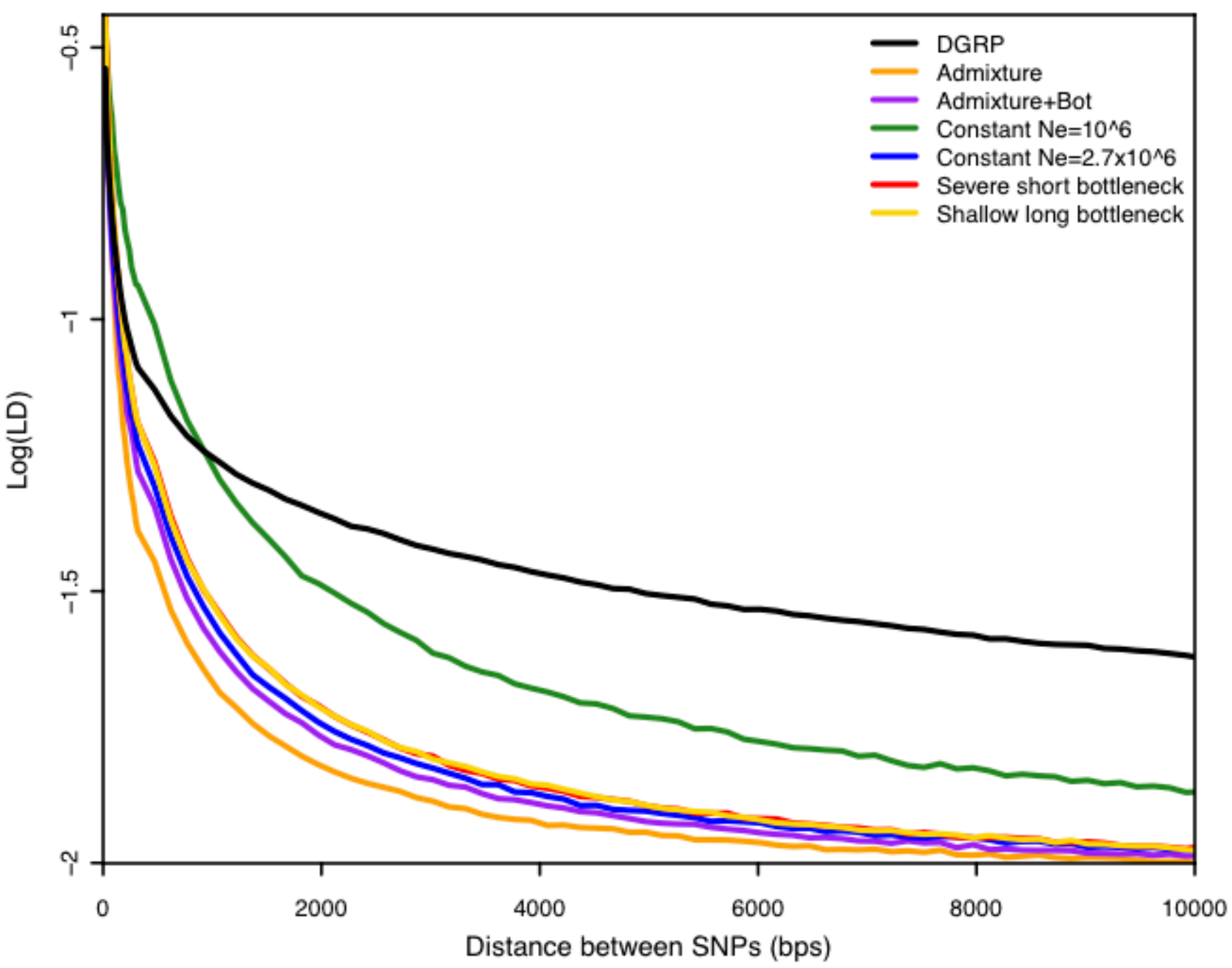

**Figure 3: Number of adaptive haplotypes in sweeps of varying softness.** The number of origins of adaptive mutations on unique haplotype backgrounds was measured in simulated sweeps of varying softness arising from (A) *de novo* mutations with $\theta_A$ values ranging from $10^{-2}$ to $10^{2}$ and (D) SGV with starting frequencies ranging from $10^{-6}$ to $10^{-1}$. Sweeps were simulated under a constant $N_e = 10^6$ demographic model with a recombination rate of $5\times10^{-7}$ cM/bp, selection strength of $s = 0.01$, partial frequency of $PF$=1 and 0.5, and in sample sizes of 145 individuals. 1000 simulations were averaged for each data point. Additionally, shown are sample haplotype frequency spectra for incomplete (B) and complete (C) sweeps arising from *de novo* mutations as well as incomplete (E) and complete (F) sweeps arising from SGV. (G) shows haplotype frequency spectra for a random simulation under the six neutral models considered in this paper in the following order: admixture, admixture with bottleneck, constant $N_e = 10^6$, constant $N_e = 2.7\text{x}10^6$, severe short bottleneck, and shallow long bottleneck models. The height of the first bar (light blue) in each frequency spectrum indicates the frequency of the most prevalent haplotype in the sample of 145 individuals, and heights of subsequent colored bars indicate the frequency of the second, third, and so on most frequent haplotypes in a sample. Grey bars indicate singletons. Sweeps generated with a low $\theta_A$ or low starting partial frequency of the adaptive allele have one frequent haplotype in the sample and look hard. In contrast, sweeps look increasingly soft as the $\theta_A$ or starting partial frequency increase and there are multiple frequent haplotypes in the sample.

**Sweeps from *de novo* mutations**

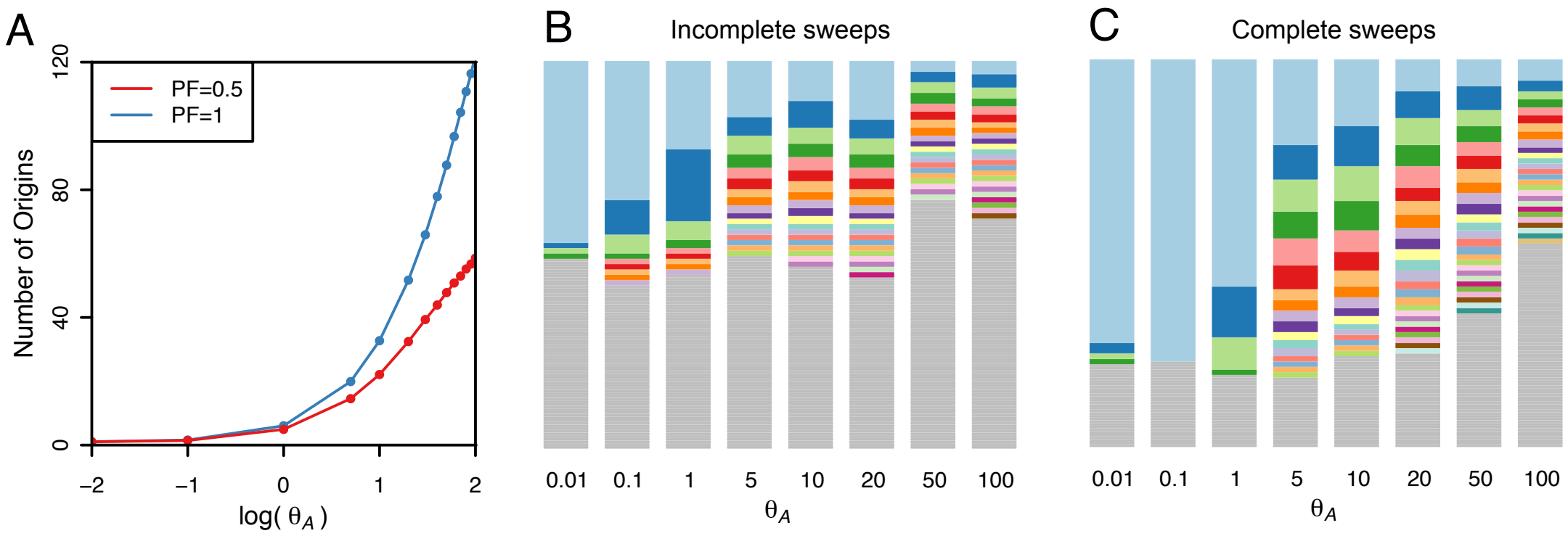

**Sweeps from standing genetic variation**

**Neutrality**

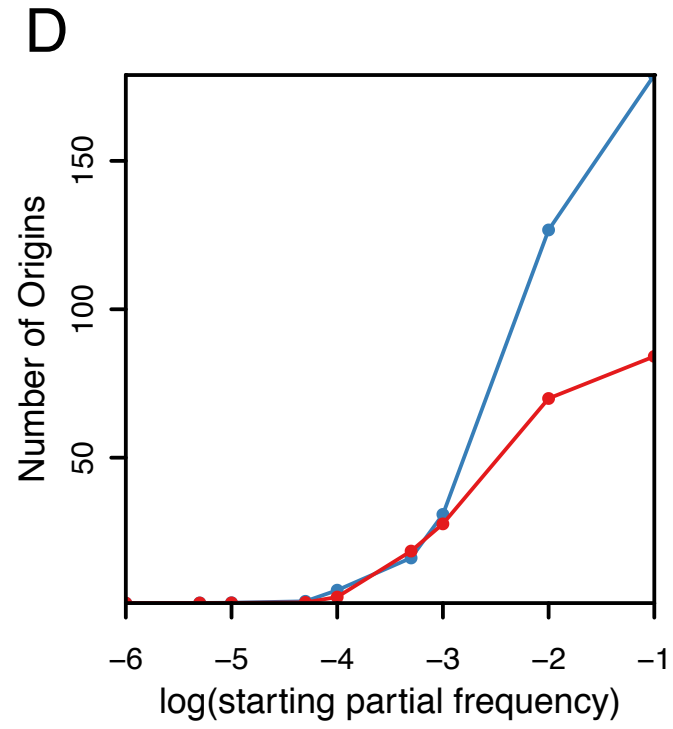

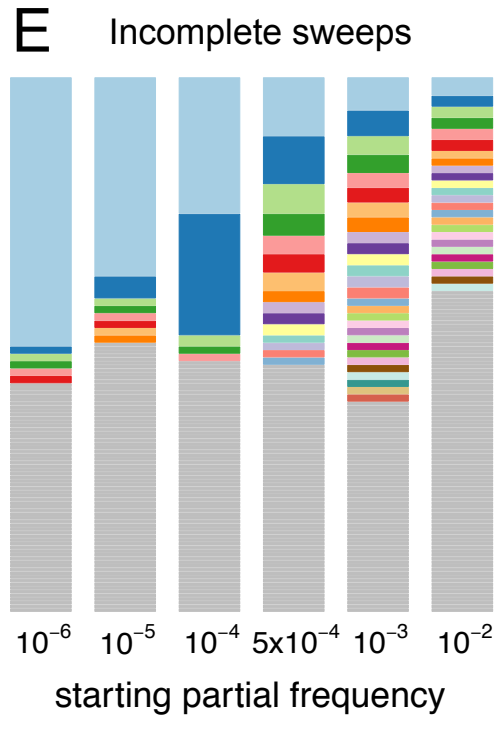

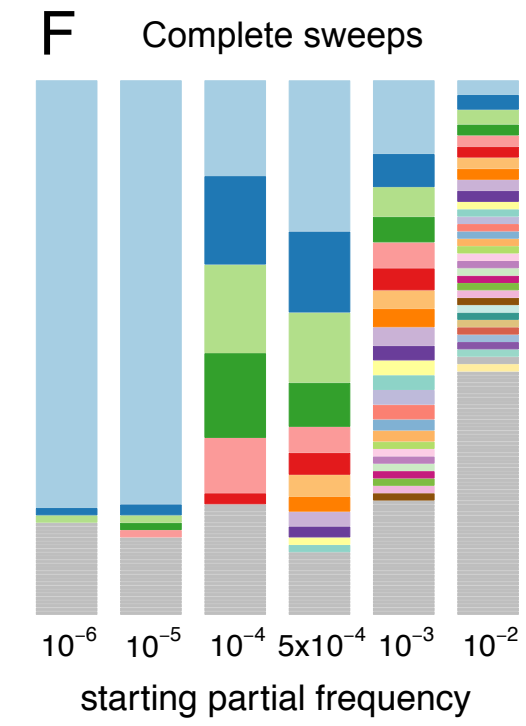

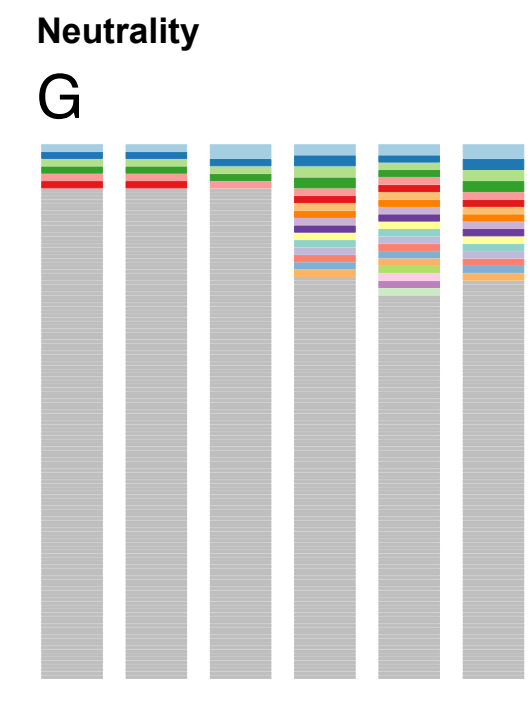

**Figure 4: Haplotype homozygosity statistics.** Depicted are haplotype frequencies for hard (red) and soft (blue) sweeps. The top row shows incomplete hard sweeps with one prevalent haplotype present in the population at frequency $p_1$, and all other haplotypes present as singletons. The bottom row shows incomplete soft sweeps with one primary haplotype with frequency $p_1$ and a second, less abundant haplotype at frequency $p_2$, with the remaining haplotypes present as singletons. Each edge of the square represents haplotype frequencies ranging from 0 to 1. (A) H1 is the sum of the squares of frequencies of each haplotype in a sample. The total H1 value corresponds to the total colored area. Hard sweeps are expected to have a higher H1 value than soft sweeps. (B) In H12, the first and second most abundant haplotype frequencies in a sample are combined into a single combined haplotype frequency and then homozygosity is recalculated using this revised haplotype frequency distribution. By combining the first and second most abundant haplotypes into a single group, H12 should have more similar power to detect hard and soft sweeps than H1. (C) H2 is the haplotype homozygosity calculated after excluding the most abundant haplotype. H2 is expected to be larger for soft sweeps than for hard sweeps. We ultimately use the ratio H2/H1 to differentiate between hard and soft sweeps as we expect this ratio to have even greater discriminatory power than H2 alone.

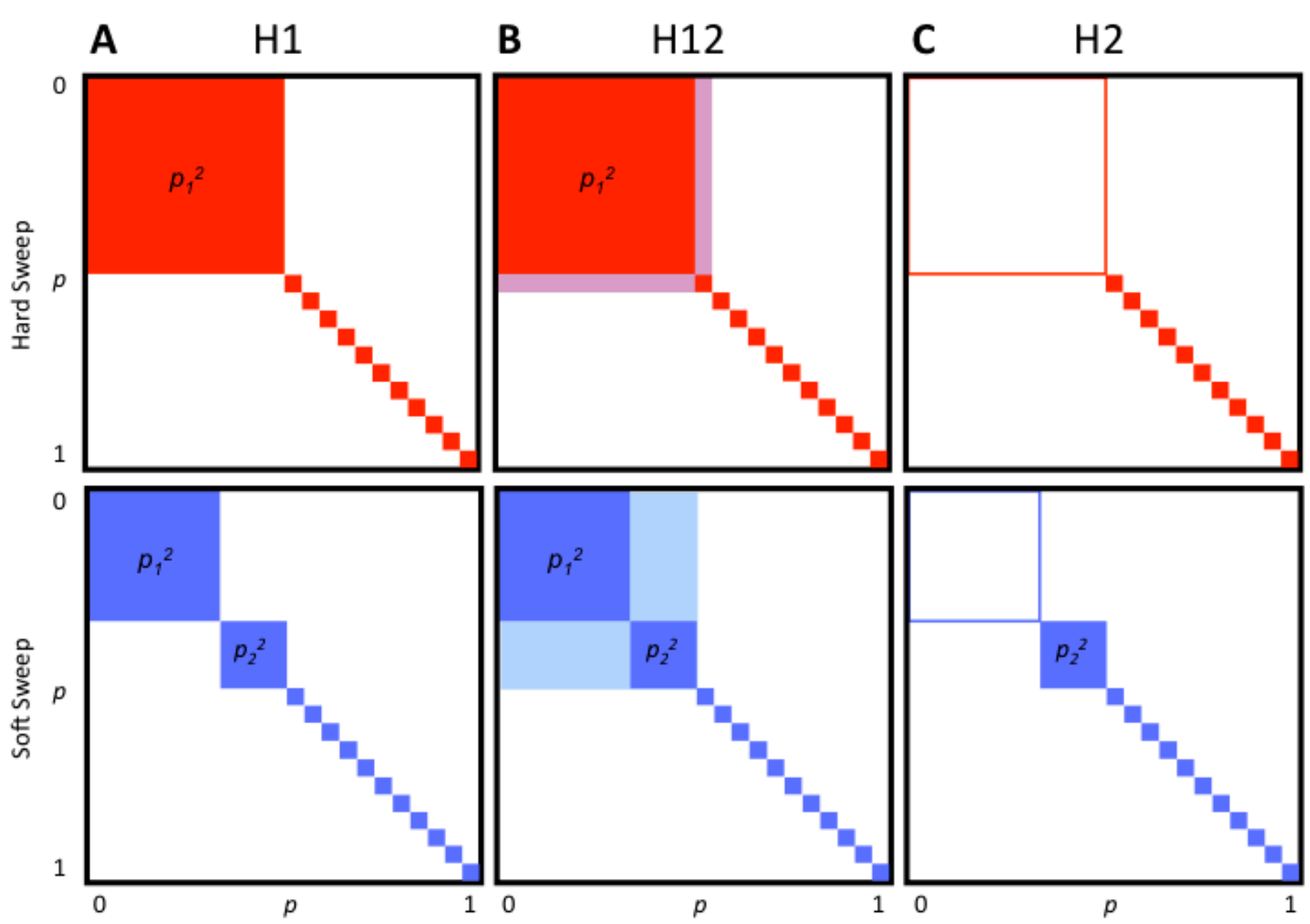

**Figure 5: H12 values in sweeps of varying softness.** H12 values were measured in simulated sweeps arising from (A) *de novo* mutations with $\theta_A$ values ranging from $10^{-2}$ to $10^{2}$ and (B) SGV with starting frequencies ranging from $10^{-6}$ to $10^{-1}$. Sweeps were simulated under a constant $N_e = 10^6$ demographic model with a recombination rate of $5\times10^{-7}$ cM/bp, selection strength of $s = 0.01$, partial frequencies $PF = 1$ and 0.5, and in samples of 145 individuals. Each data point was averaged over 1000 simulations. H12 values rapidly decline as the softness of a sweep increases and as the ending partial frequency of the sweep decreases. In (C) and (D), $s$ was varied while keeping $PF$ constant at 0.5 for sweeps from *de novo* mutations and SGV, respectively. H12 values increase as $s$ increases, though for very weak $s$ we observe a 'hardening' of sweeps where fewer adaptive alleles reach establishment frequency. In (E) and (F), the time since selection ended ($T_E$) was varied for incomplete ($PF$=0.5) and complete ($PF$=1) sweeps respectively while keeping $s$ constant at 0.01. As the age of a sweep increases, sweep signatures decay and H12 loses power.

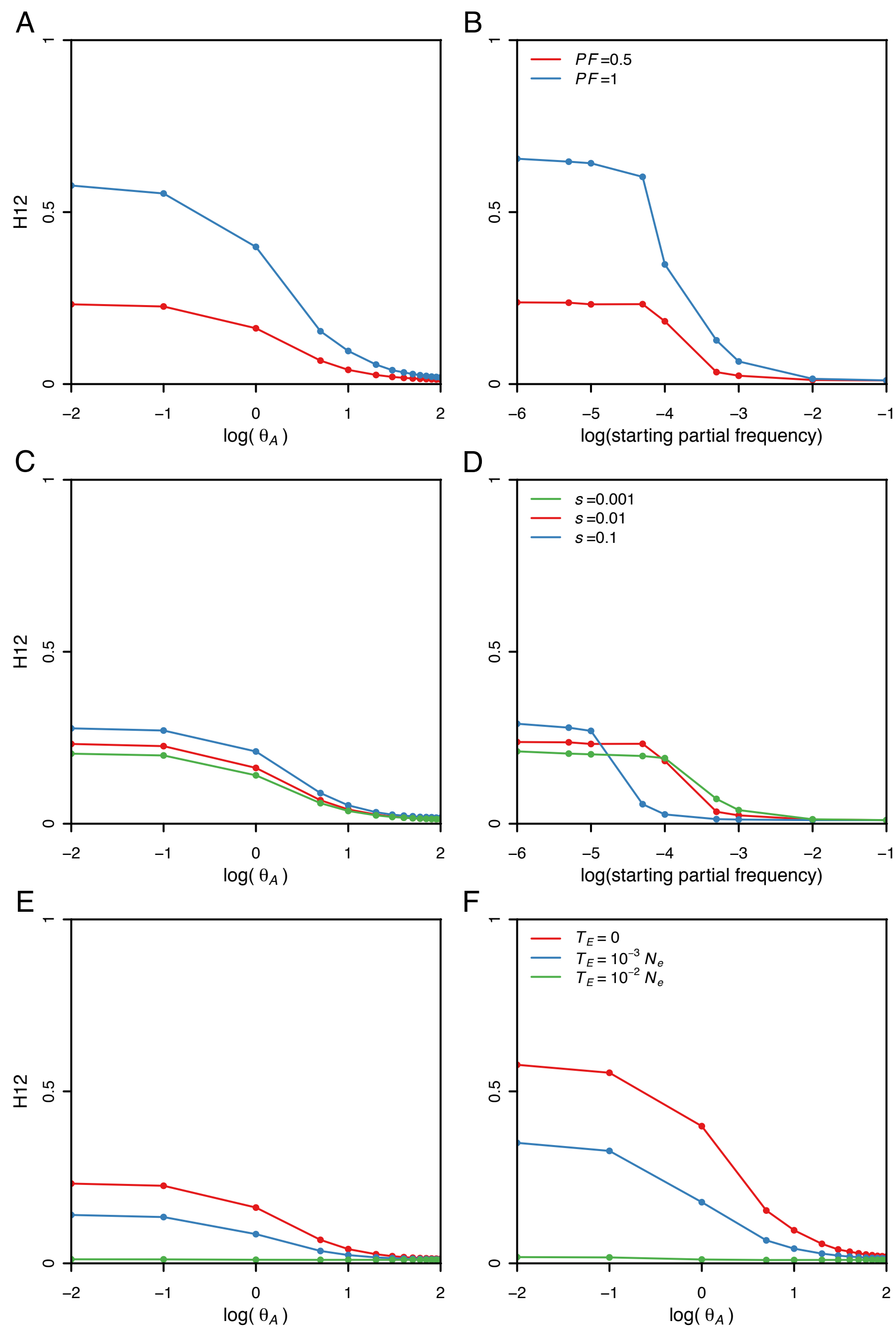

**Figure 6: Power analysis of H12 and *iHS* under different sweep scenarios.** The plots show ROC curves for H12 and *iHS* under various sweep scenarios with the specified selection coefficients ($s$), and the time of the end of selection ($T_E$) in units of $4N_e$ generations. In all scenarios, the partial frequency of the sweeps was 0.5. False positive rates (FPR) were calculated

by counting the number of neutral simulations that were misclassified as sweeps under a specific cutoff. True positive rates (TPR) were calculated by counting the number of simulations correctly identified as sweeps under the same cutoff. Hard and soft sweeps were simulated from *de novo* mutations with $\theta_A = 0.01$ and 10, respectively, under a constant effective population size of $N_e = 10^6$, a neutral mutation rate of $10^{-9}$ bp/gen, and a recombination rate of $5\times10^{-7}$ cM/bp. A total of 5000 simulations were conducted for each evolutionary scenario. H12 performs well in identifying recent and strong selective sweeps, and is more powerful than *iHS* in identifying soft sweeps.

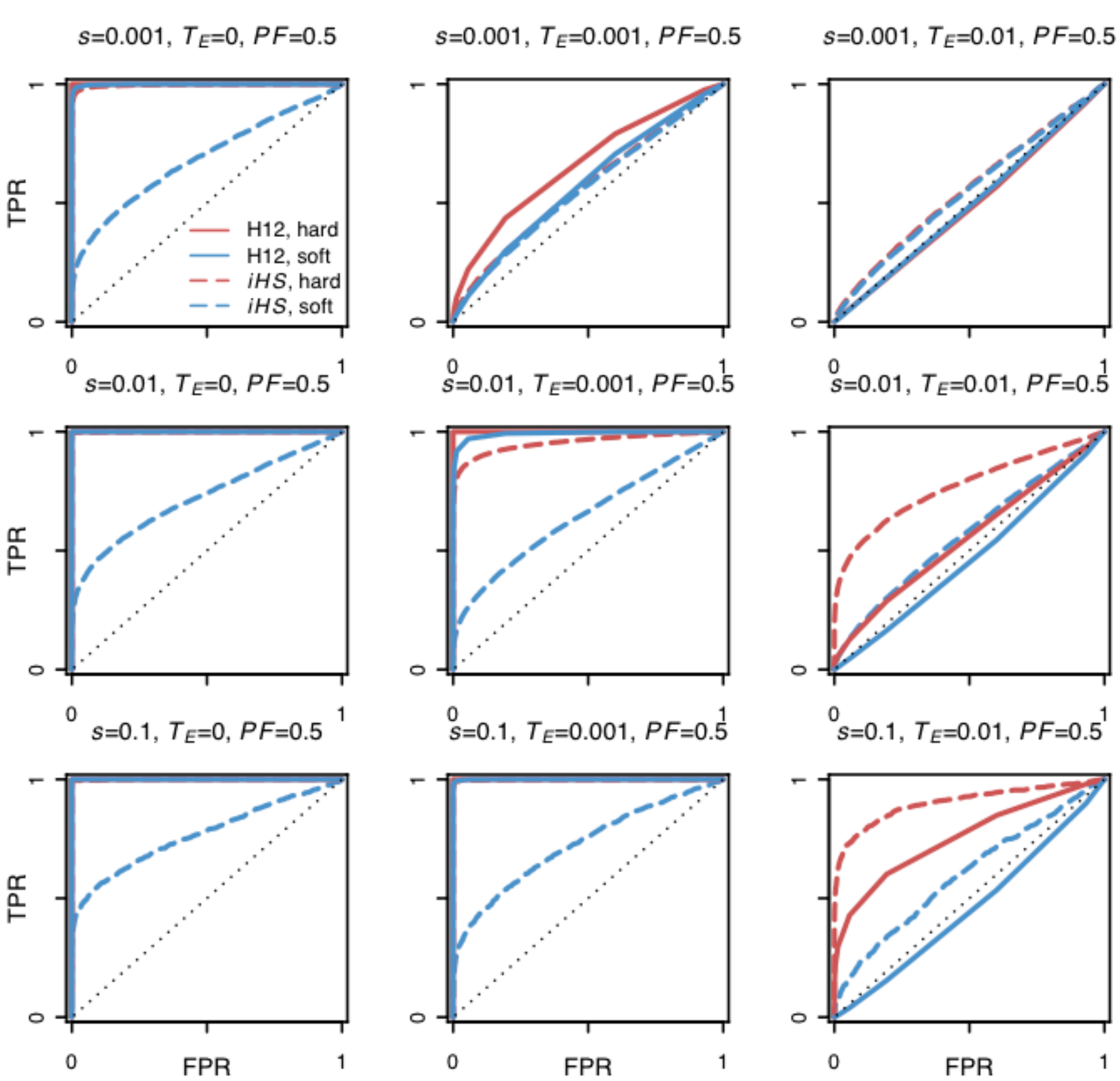

**Figure 7: Elevated H12 values and long-range LD in DGRP data.** (A) Genome-wide H12 values in DGRP data are elevated as compared to expectations under any neutral demographic model tested. Plotted are H12 values for DGRP data reported in analysis windows with $\varrho \geq 5\times10^{-7}$ cM/bp. Red dots overlaid on the distribution of H12 values for DGRP data correspond to the highest H12 values in outlier peaks of the DGRP scan at the 50 top peaks depicted in Figure 8A. Note that most of the points in the tail of the H12 values calculated in DGRP data are part of the top 50 peaks as well. Neutral demographic simulations were generated with $\varrho = 5\times10^{-7}$ cM/bp. Plotted are the result of approximately $1.3\text{x}10^{5}$ simulations under each neutral demographic model, representing ten times the number of analysis windows in DGRP data.

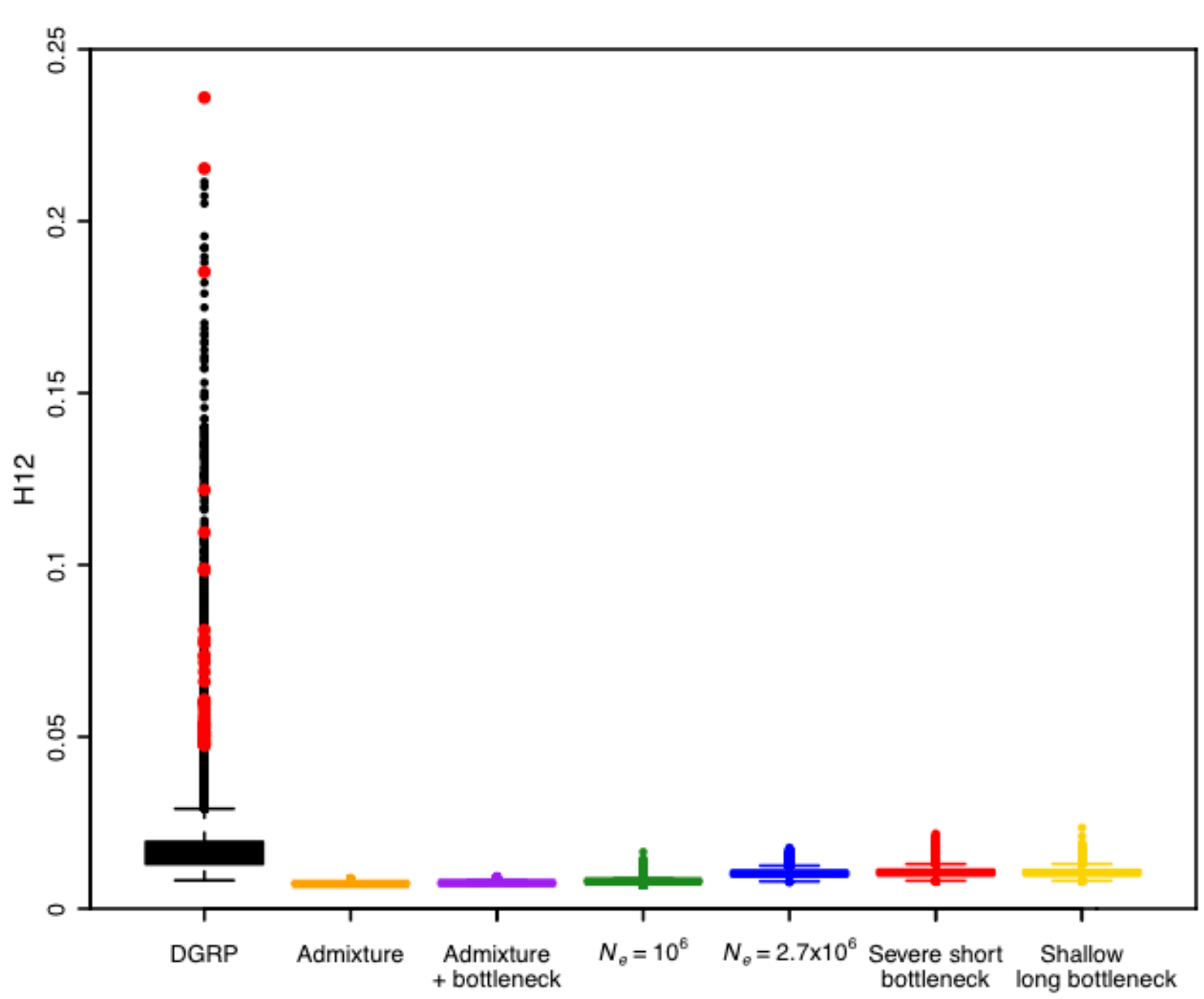

**Figure 8: H12 and *iHS* scan in DGRP data along the four autosomal arms.** (A) H12 scan. Each data point represents the H12 value calculated over an analysis window of size 400 SNPs centered at the particular genomic position. Grey points indicate regions in the genome with recombination rates lower than $5\times10^{-7}$ cM/bp we excluded from our analysis. The orange line represents the 1-per-genome FDR line calculated under a neutral demographic model with a constant population size of $10^{6}$ and a recombination rate of $5\times10^{-7}$ cM/bp. Red and blue points highlight the top 50 H12 peaks in the DGRP data relative to the 1-per-genome FDR line. Red points indicate the peaks that overlap the top 10% of 100Kb windows with an enrichment of SNPs with $|iHS| > 2$ in (B). We identify three well-characterized cases of selection in *D. melanogaster* at *Ace*, *CHKov1*, and *Cyp6g1* as the three highest peaks. (B) *iHS* scan. Plotted are the number of SNPs in 100Kb windows with standardized *iHS* values ($|iHS|$) > 2. Highlighted in red and blue are the top 10%100Kb windows (total of 95 windows). Red points correspond to those windows that overlap the top 50 peaks in the H12 scan. The positive controls, *Ace*, *CHKov1*, and *Cyp6g1* are all among the top 10%.

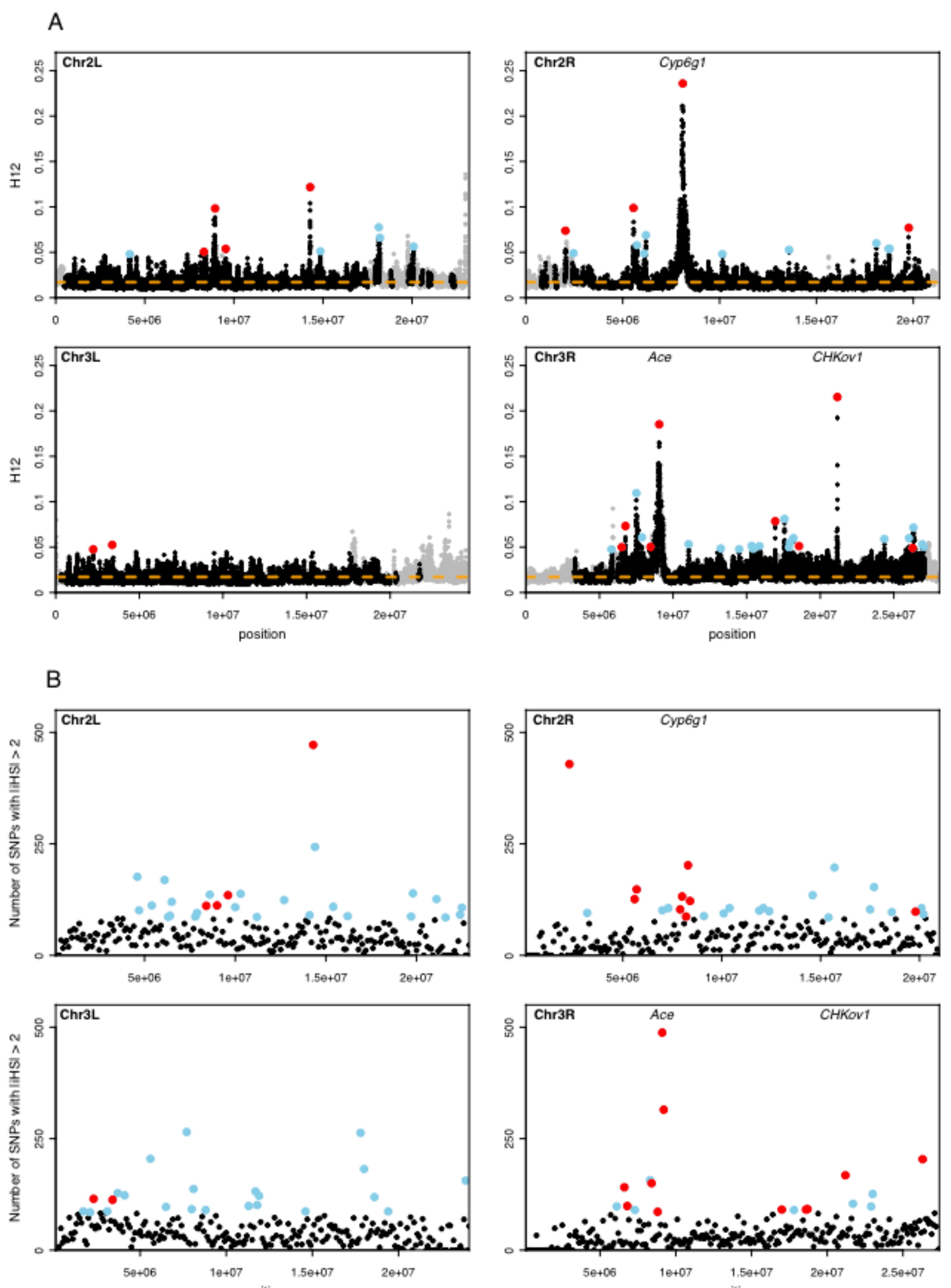

**Figure 9: Haplotype frequency spectra for the top 10 peaks and extreme outliers under neutral demographic scenarios.** (A) Haplotype frequency spectra for the top 10 peaks in the DGRP scan in order from highest to lowest H12 value. For each peak, the frequency spectrum corresponding to the analysis window with the highest H12 value is plotted, which should be the "hardest" part of any given peak. As can be seen, at all peaks there are multiple haplotypes present at high frequency, compatible with signatures of soft sweeps shown in Figure 5. In none of the cases is there one single haplotype present at high frequency, as would be expected in a hard sweep. (B) In contrast, the haplotype frequency spectra corresponding to the extreme outliers under the six neutral demographic scenarios have critical $H12_0$ values that are significantly lower than the H12 values at the top 10 peaks. The order of the neutral demographic models whose spectra are shown are as follows: admixture, admixture with a bottleneck, constant $N_e$=$10^6$, constant $N_e$=2.7x$10^6$, severe short bottleneck, and shallow long bottleneck.

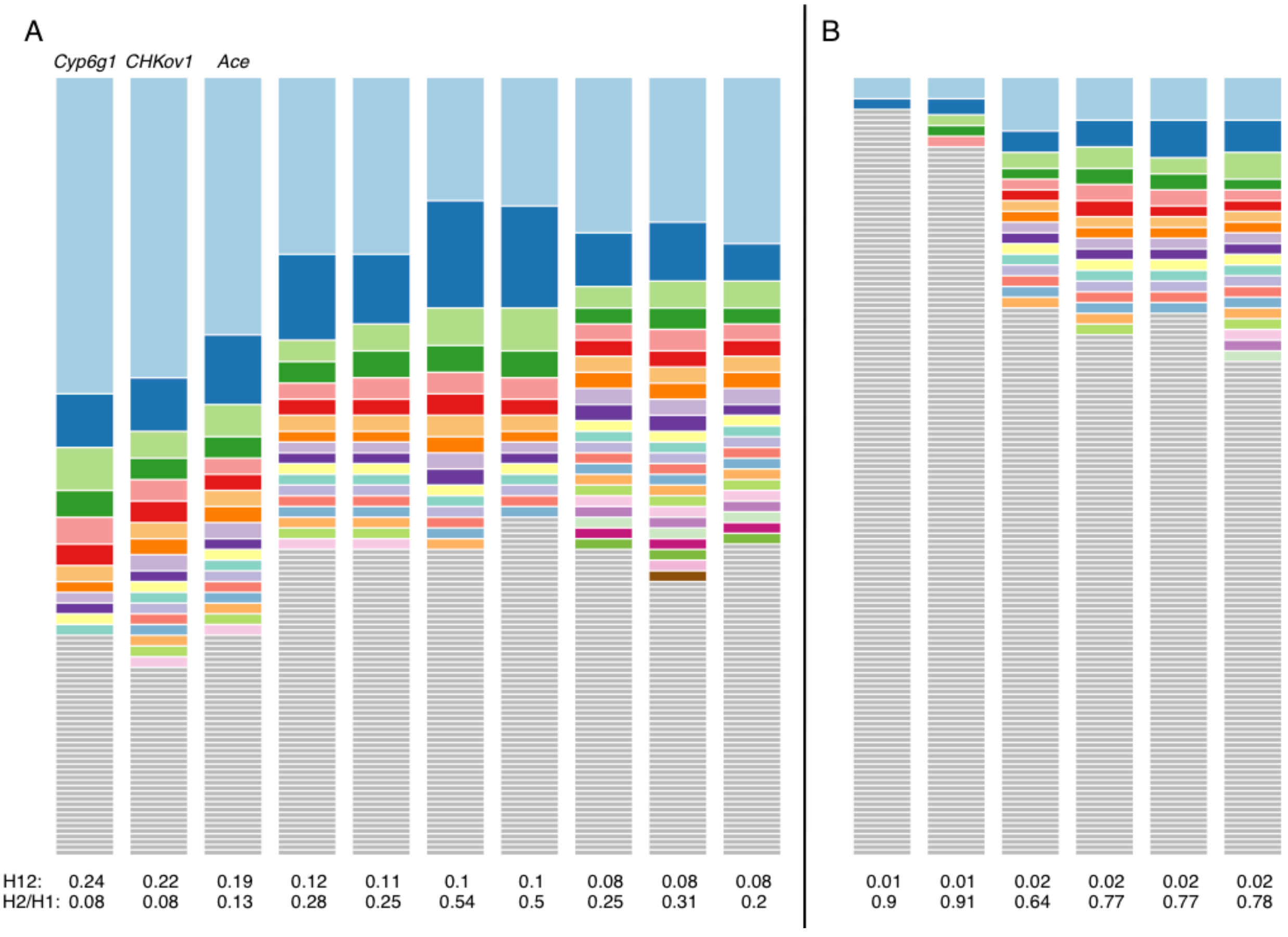

**Figure 10: H2/H1 values measured in sweeps of varying softness.** Similar to Figure 5, H2/H1 values were measured in simulated sweeps arising from (A) *de novo* mutations with $\theta_A$ values ranging from $10^{-2}$ to $10^{2}$ and (B) SGV with starting frequencies ranging from $10^{-6}$ to $10^{-1}$. Sweeps were simulated under a constant $N_e = 10^6$ demographic model with a recombination rate of $5\times10^{-7}$ cM/bp, selection strength of $s = 0.01$, partial frequencies $PF = 1$ and 0.5, and in samples of 145 individuals. Each data point was averaged over 1000 simulations. H2/H1 values rapidly increase with increasing softness of a sweep, but do not depend strongly on the ending partial frequency of the sweep (*PF)*. In (C) and (D), *s* was varied while keeping *PF* constant at 0.5 for sweeps from *de novo* mutations and SGV, respectively. H2/H1 values increase as *s* increases in the case of sweeps from SGV reflecting a hardening of sweeps with smaller *s*. In (E) and (F), the time since selection ended ($T_E$) was varied for incomplete ($PF$=0.5) and complete ($PF$=1) sweeps respectively while keeping *s* constant at 0.01. As the age of a sweep increases, the sweep signature decays and H2/H1 approaches one.

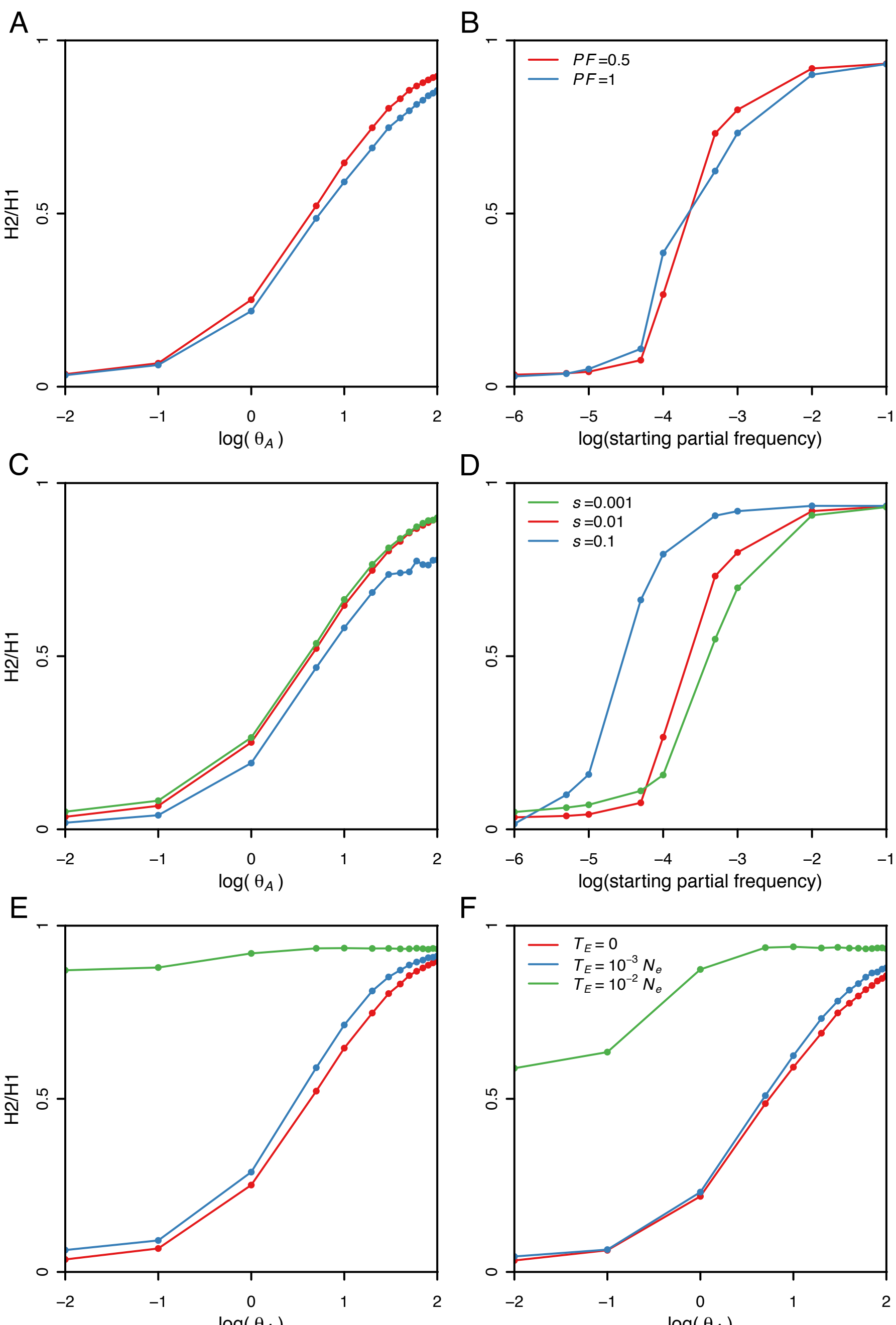

**Figure 11: Range of H12 and H2/H1 values expected for hard and soft sweeps.**

Bayes factors (BFs) were calculated for a grid of H12 and H2/H1 values to demonstrate the range of H12 and H2/H1 values expected under hard versus soft sweeps. Each panel shows the results for a specific evolutionary scenario defined by the underlying demographic model, the $\theta_A$ value used for simulating soft sweeps, and the recombination rate as specified below. BFs were calculated by taking the ratio of the number of soft sweep versus hard sweep simulations that were within a Euclidean distance of 10% of a given pair of H12 and H2/H1 values. Red portions of the grid represent H12 and H2/H1 values that are more easily generated by hard sweeps, while grey portions represent regions of space more easily generated under soft sweeps. Each panel presents the results from one million hard and soft sweep simulations. Hard sweeps were always generated with $\theta_A = 0.01$. (A), (B), and (C) compare the range of BFs obtained when soft sweeps are generated under $\theta_A$ = 5, 10, and 50, keeping the recombination rate, $\rho$, constant at $5\times10^{-7}$ cM/bp. (A), (D), and (E) compare the range of BFs obtained when $\rho$ is varied from $5\times10^{-7}$, $10^{-7}$, and $10^{-6}$, keeping the $\theta_A$ constant at 10. (A) and (F) compare the range of BFs generated under the constant $N_e = 10^6$ and admixture demographic models for $\theta_A = 10$ and $\rho = 5\times10^{-7}$ cM/bp. When H12 values are smaller than 0.05, there is little evidence for a sweep, and most BFs are smaller than one. As H12 values become larger, virtually all sweeps with H2/H1 values > 0.05 are soft. The 50 yellow points show the observed H12 and H2/H1 values for the top 50 peaks in the DGRP scan. All sweep candidates have H12 and H2/H1 values that are more easily generated by soft sweeps than hard sweeps in most scenarios.

**A**: Constant $N_e = 10^6$, soft sweeps simulated with $\theta_A = 10$, $\rho = 5\times10^{-7}$ cM/bp

**B**: Constant $N_e = 10^6$, soft sweeps simulated with $\theta_A = 5$, $\rho = 5\times10^{-7}$ cM/bp

**C**: Constant $N_e = 10^6$, soft sweeps simulated with $\theta_A = 50$, $\rho = 5\times10^{-7}$ cM/bp

**D:** Constant $N_e = 10^6$, soft sweeps simulated with $\theta_A = 10$, $\rho = 10^{-7}$ cM/bp

**E:** Constant $N_e = 10^6$, soft sweeps simulated with $\theta_A = 10$, $\rho = 10^{-6}$ cM/bp

**F:** Admixture, soft sweeps simulated with $\theta_A = 10$, $\rho = 5\times10^{-7}$ cM/bp

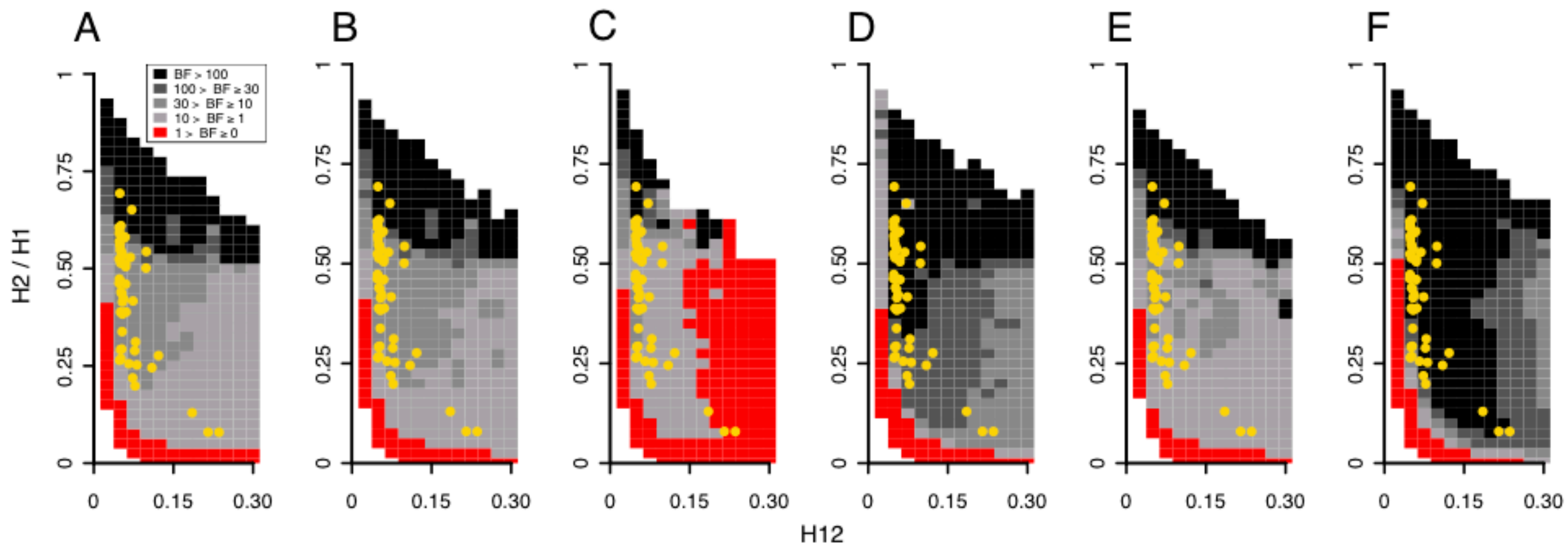

**Table 1: FDR critical $H12_o$ values for different demographic models and recombination rates.** For our genomic scan we chose to use the 1-per-genome FDR value calculated under the constant $N_e = 10^6$ model with a recombination rate of $5 \times 10^{-7}$ cM/bp. Note that most $H12_o$ values are similar to the genome-wide median H12 value of 0.0155.

| **Demographic model** | **$\rho = 10^{-7}$ cM/bp** | **$\rho = 5 \times 10^{-7}$ cM/bp** | **$\rho = 10^{-6}$ cM/bp** |
|---|---|---|---|
| Admixture | 0.0084 | 0.0083 | 0.0083 |
| Admixture and bottleneck | 0.0141 | 0.0092 | 0.0085 |
| Constant $N_e = 10^6$ | 0.0391 | 0.0171 | 0.0126 |
| Constant $N_e = 2.7 \times 10^6$ | 0.0383 | 0.0168 | 0.0133 |
| Severe short bottleneck | 0.0450 | 0.0187 | 0.0131 |
| Shallow long bottleneck | 0.0398 | 0.0181 | 0.0083 |

## SUPPLEMENTARY TEXT

### Calculation of the 1-per-genome FDR critical value of H12 $_o$.

We calculated the critical values, H12$_o$, six neutral models for three different recombination rates, $\rho$ = $10^{-7}$, $5\times10^{-7}$, and $10^{-6}$ cM/bp, based on a 1-per-genome false discovery rate (FDR) criterion. Our test rejects neutrality in favor of a selective sweep when H12 > H12$_o$ . The critical values H12$_o$ for rejecting neutrality with a given recombination rate, $\rho_0$, are conservative for genomic regions with recombination rates $\rho > \rho_0$ (Table 1). Note that H12$_o$ values obtained under models with the lowest recombination rate ($\rho$ = $10^{-7}$ cM/bp) are substantially higher than H12$_o$ values calculated under models with recombination rates even modestly higher than $10^{-7}$ cM/bp. Therefore, H12$_o$ values calculated under low recombination rates may be too conservative for most genomic regions. Hence, we used the H12$_o$ value obtained from regions with an intermediate $\rho$ = $5\times10^{-7}$ cM/bp, filtering out all regions with a recombination rate lower than $5\times10^{-7}$ cM/bp from the data.

### Robustness of the H12 scan

To ensure that the H12 peaks identified in our genomic scan are robust to any peculiarities of the DGRP data set such as inversions, unaccounted substructure within the data, or sequencing quality, we performed a number of tests: The individual strains of the DGRP data set contain a number of inversions, seven of which are shared across multiple strains (Table S4A) (The locations of inversion breakpoints were identified by Spencer Koury, personal communication). One possibility is that elevated peaks of homozygosity could result from inversions suppressing recombination. To test for this possibility, we performed a binomial two-sided test for enrichment of the top 50 peaks in regions with inversions versus a model of a uniform distribution of the peaks genome-wide. We found no significant enrichment in any inversion except for an inversion on chromosome 3R, In(3R)K (P-value=6.44E-06) (Table S4A). We further performed a chi-square test for a correlation between members of haplotype groups in each peak and haplotypes potentially linked to an inversion on the same chromosome, as inversions have been shown to affect polymorphisms chromosome-wide [54]. We did not find any enrichment for strains bearing inversions in any single haplotype cluster group for the top 50 peaks (Chi-Square test, Table S4B), suggesting that the enrichment of peaks in the In(3R)K inversion cannot be attributable to inversions. Finally, even after removing regions of the genome overlapping major cosmopolitan inversions, there continues to be an elevation and long tail of H12 values in DGRP data relative to expectations under any neutral demographic model (Figure S6).

During our analysis of the DGRP data set, two new data sets based on the same North Carolina population of flies became available: the Drosophila Population Genomics Project (DPGP) data set, which consists of 40 of the original 162 inbred lines in the DGRP data set, and version 2 of the DGRP version data set, comprised of 205 lines including the original 162 lines.

Given the shallower sample depth, we scanned the DPGP data set with a window size of 100 SNPs and found that 16 peaks of the top 50 in the DPGP scan overlap 13 of the top 50 unique peaks in the DPGP scan (Figure S7A). Ten of these overlapping peaks are among the top 15 peaks in the DGRP scan. We define an overlap of two peaks as an intersection of the edge coordinates of the first and last windows in the two peaks.

We repeated the analysis in the DGRP version 2 data set as well. In the DGRP data set, there are at least five pairs of strains with genome-wide identity by descent (IBD) values > 50% suggesting twin or sibling relationships [42], and three of these complete pairs were among our data set of 145 strains. Since related strains can increase homozygosity, in our new DGRP v2 scan, we removed one of the members of each closely related pair to ensure that the top 50 H12 peaks are robust to any homozygosity contributed by related pairs of flies. In addition, we removed strains with the most missing data, and down sampled to 145 lines to match the number of strains in the original scan. Forty of the top 50 DGRPv2peaks overlapped 34 unique peaks among the top 50 peaks in the DGRP scan (Figure S7B). Since related pairs can increase homozygosity at most by $(2/145)^2 = 0.00019$, we did not exclude these lines from the final analysis of the DGRP data.

We scanned the remaining 63 strains that were non-overlapping with the original 145 strains to determine if we could recover the peaks in a completely independent data set, and observed that 12 peaks among the top 50 peaks in this scan overlap 11 unique peaks among the top 50 peaks identified in the DGRP data set (Figure S7C).

Finally, we sub-sampled the DGRP data set to 40 strains 10 times and plotted the resulting distributions of H12 values (Figure S8). In comparison to H12 distributions observed in the six tested neutral demographic models also sampled at 40 strains, there is an elevation and long tail of genome-wide H12 values, indicating that the elevation in homozygosity observed in the DGRP data are population-wide and specific to a sub population.

**Estimates of $\theta_A$ for the top 50 peaks**

The monotonic relationship between the softness of a sweep and both H12 and H2/H1 over the interval ($0.01 < \theta_A < 100$) in Figures 5 and 10 suggests that these two statistics are informative for the purpose of inferring the softness of a sweep. Here, we estimate the softness of a sweep by varying the parameter $\theta_A$. We developed a Bayesian approach for inferring $\theta_A$ by sampling the posterior distribution of $\theta_A$ conditional on the observed values $H12_{obs}$ and $H2_{obs}/H1_{obs}$ from a candidate sweep. Given that sampling this true posterior distribution is computationally intractable, we used approximate Bayesian computation (ABC) for our inference procedure. Specifically, we drew $\theta_A$ values from a prior distribution, simulated a large data set under each $\theta_A$ value, and then kept 1000 parameter values which produce sweeps with H12 and H2/H1 values close to the observed values $H12_{obs}$ and $H2_{obs}/H1_{obs}$ from the candidate sweep (differences <10% for each statistic). From these posterior distributions, we inferred the maximum a posteriori ($\theta_A^{MAP}$) value of the given candidate sweep to estimate its softness (Methods).

We estimated the softness of the top 50 peaks detected in our H12 scan in Figure 8A by inferring the $\theta_A^{MAP}$ value that generates haplotype structure best resembling the spectra observed for each peak using the above ABC procedure. We first considered the $N_e = 10^6$ demographic model and uniform prior distributions for all other parameters: The adaptive mutation rate $\theta_A$ took values on [0,100], the selection coefficient ($s$) on [0,1], the ending partial frequency of the sweep ($PF$) on [0,1], the time at which selection ended ($T_E$) on $[0,0.001]\times 4N_e$, and the recombination rate ($\rho$) on an interval containing the observed recombination rate at each peak (see Methods).

The posterior distributions of $\theta_A$ and the estimates of $\theta_A^{MAP}$ for the top nine peaks obtained by our procedure are shown in Figure S10A. The distribution of $\theta_A^{MAP}$ values for all top 50 peaks is shown in Figure S10B. Table S4 lists all $\theta_A^{MAP}$ values and their 95% confidence intervals. The minimum $\theta_A^{MAP}$ value among all 50 top peaks is $\theta_A^{MAP} = 6.8$, which is obtained for the peak centered at *Cyp6g1*.

We also estimated $\theta_A^{MAP}$ for our top 50 peaks under the admixture model proposed by Duchen *et al*. [32] to determine the effect of admixture on our estimates (Methods). Figure S10A shows the comparison of the posterior distributions of $\theta_A$ inferred under the constant $N_e = 10^6$ and admixture models for the top nine peaks. The posterior distributions of $\theta_A$ under the admixture model tends to have a smaller variance than under the constant $N_e = 10^6$ model. Figure S10B and Table S4 show that $\theta_A^{MAP}$ estimates of the top nine peaks for the two models are similar, but slightly higher under the admixture model as compared to the constant $N_e = 10^6$ model. This suggests that the $\theta_A^{MAP}$ estimates under the constant $N_e = 10^6$ model are in fact conservative in estimating the softness of each peak.

**Figure S1: Simple bottleneck models inferred by DaDi.** The inferred parameters were the size of the final population ($N_F$), the duration of the bottleneck ($T_B$), and the time after the bottleneck ($T_F$). Investigated bottleneck sizes ranged from $N_B = 0.002$ to $N_B = 0.4$ (see Table S2). $N_B = 0.002$ represents the population size of the bottleneck inferred for European flies by Li and Stephan (2006) [55], whereas $N_B = 0.4$ represents a comparatively shallow population size reduction.

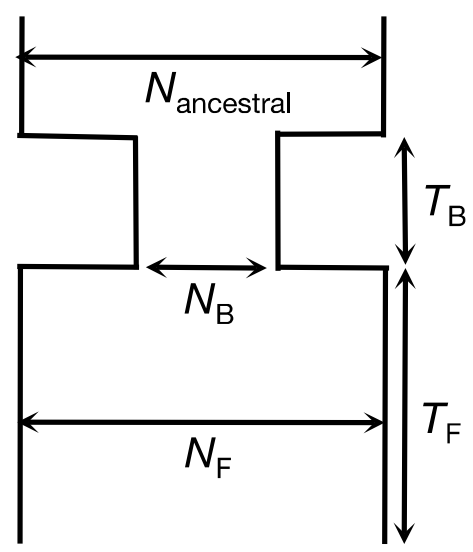

**Figure S2: Higher number of haplotypes (*K*) in under the admixture model versus the constant $N_e = 10^6$ model.** We observe a significantly higher number of unique haplotypes (*K*) in neutral simulations of admixture as compared to a constant $N_e$ scenario. Here we plot distributions of *K* in a sample of haplotypes drawn from the North American deme in the admixture model in Figure 1 and a constant $N_e = 10^6$ model. In each scenario, 1000 simulations were performed.

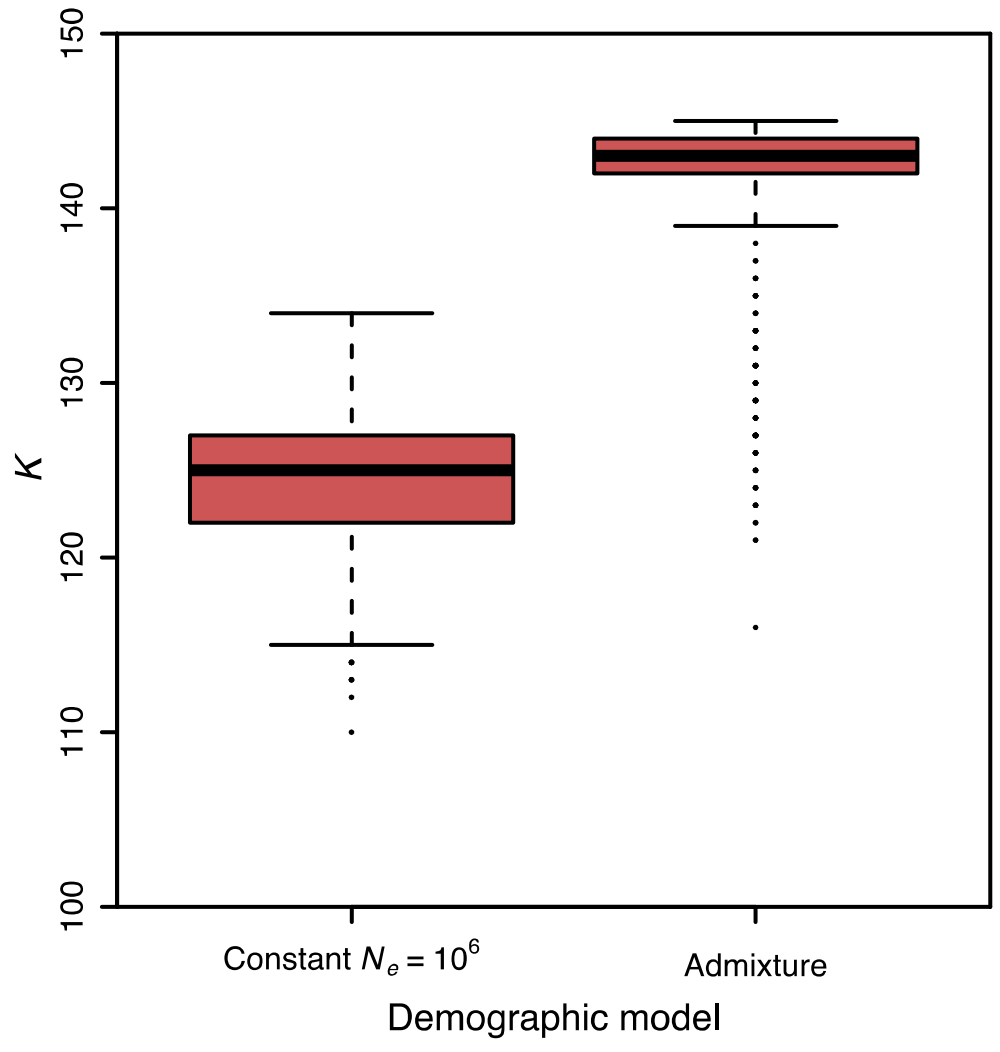

**Figure S3: H1, H12, and H123 values measured in sweeps of varying softness.** Homozygosity values were measured in simulated sweeps arising from (A) *de novo* mutations with $\theta_A$ values ranging from $10^{-2}$ to $10^{2}$ and (B) SGV with starting frequencies ranging from $10^{-6}$ to $10^{-1}$. Sweeps were simulated under a constant $N_e = 10^6$ demographic model with a recombination rate of $5\times10^{-7}$ cM/bp, selection coefficient of $s = 0.01$, and partial frequency of $PF$=0.5. Each data point was averaged over 1000 simulations. H1, H12, and H123 values all decline rapidly as the softness of a sweep increases. H12 modestly augments our ability to detect a sweep as long as the sweep is not too soft or too old. H123 has marginally better ability to detect selective sweeps as compared to H12.

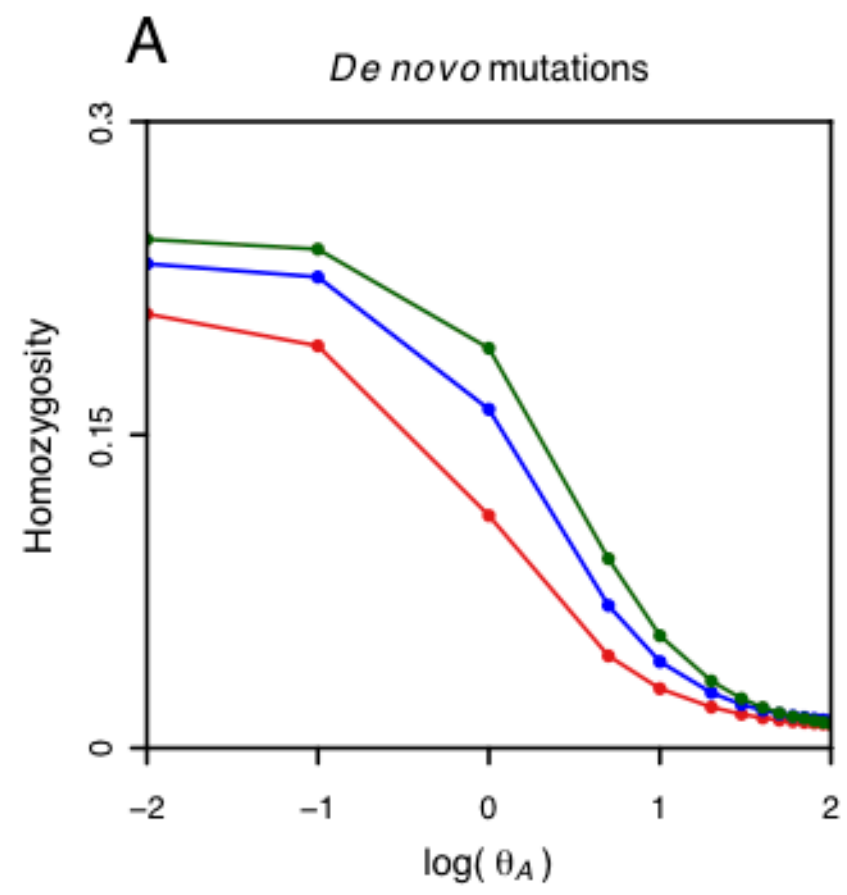

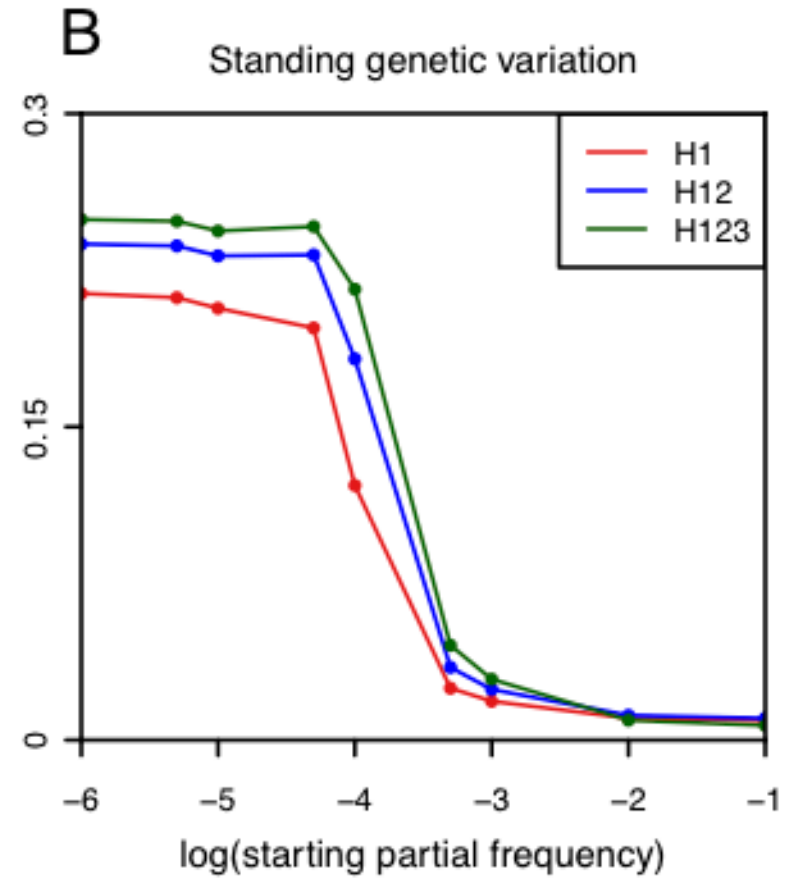

**Figure S4: Power analysis of H12 and *iHS* under different sweep scenarios.** Same as Figure 6, except ending partial frequencies of the sweeps are $PF = 0.1$ in (A) and $PF = 0.9$ in (B).

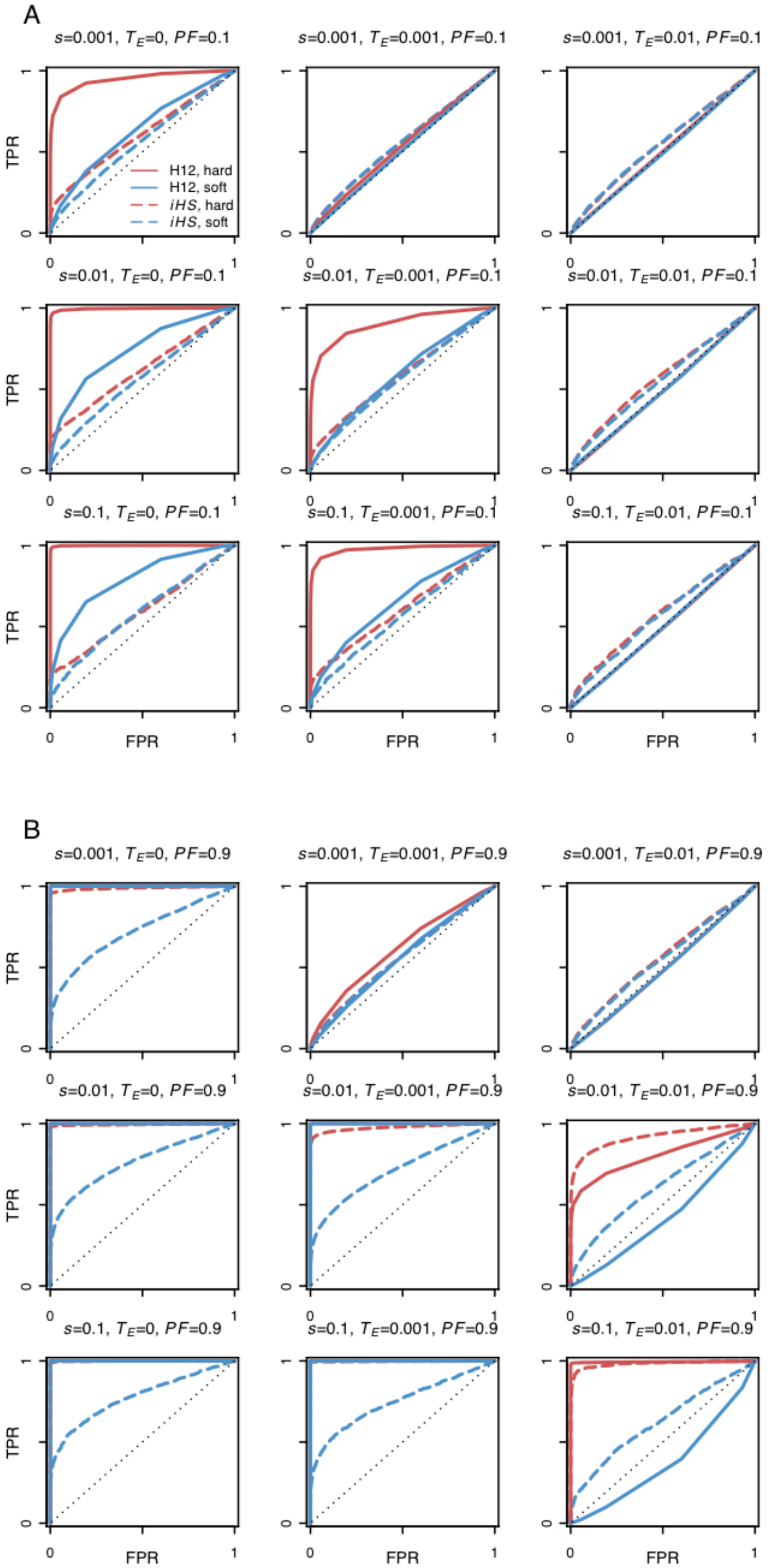

**Figure S5: Haplotype frequency spectra for the 11th-50th peaks.** Same as Figure 9, except plotted are haplotype frequency spectra for the (A)11th-30th and the (B) 31st - 50th peaks in the DGRP scan.

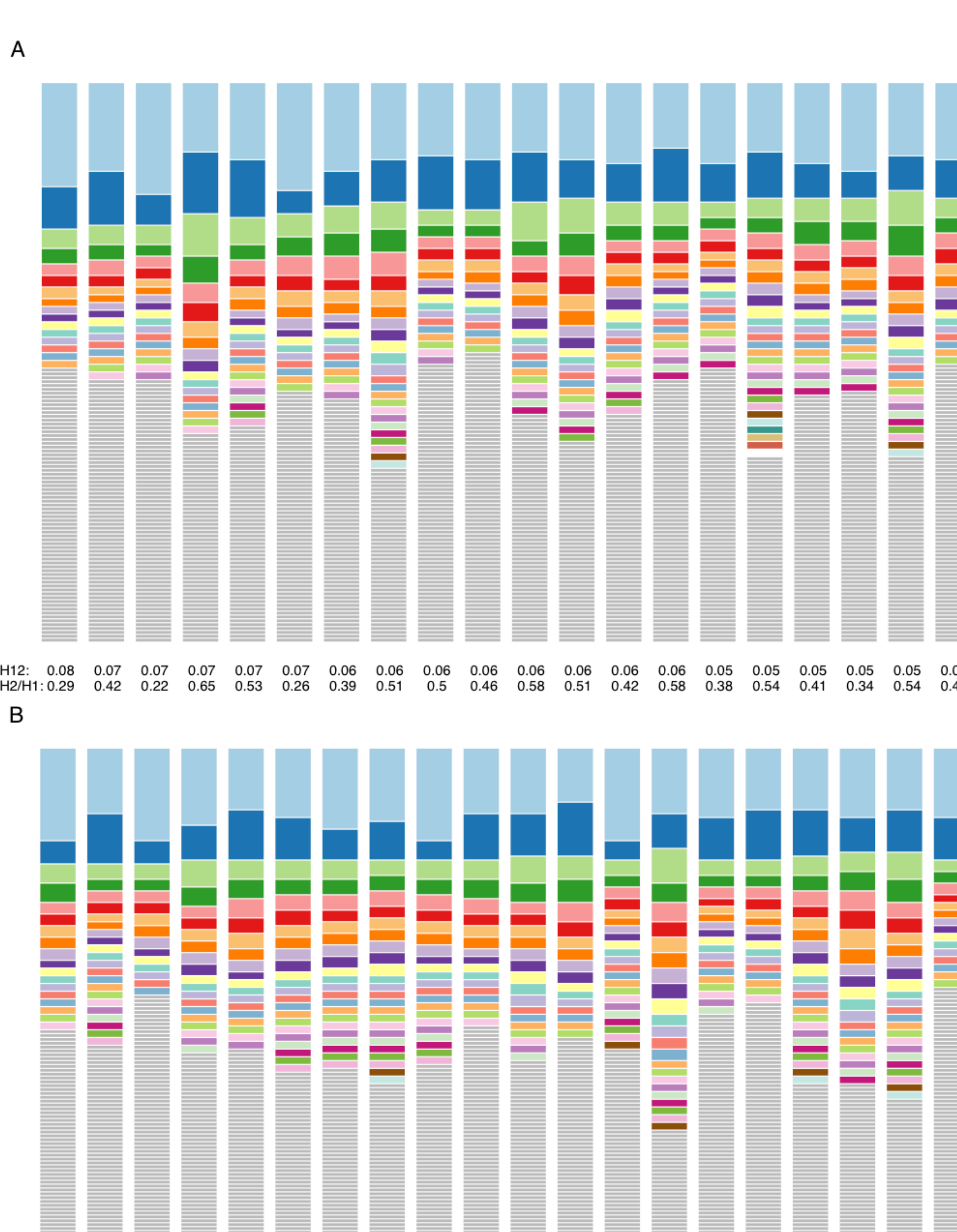

H12: 0.05 0.05 0.05 0.05 0.05 0.05 0.05 0.05 0.05 0.05 0.05 0.05 0.05 0.05 0.05 0.05 0.05 0.05 0.05 0.05
H2/H1: 0.29 0.55 0.27 0.44 0.61 0.51 0.39 0.47 0.29 0.55 0.56 0.69 0.26 0.6 0.47 0.57 0.59 0.52 0.61 0.46

**Figure S6: Elevated H12 values in DGRP data.** Similar to Figure 7, except here regions overlapping major cosmopolitan inversions are excluded from the distribution of H12 values in DGRP data. There is a long tail and elevation of H12 values in DGRP data as compared to expectations under any neutral demographic model tested.

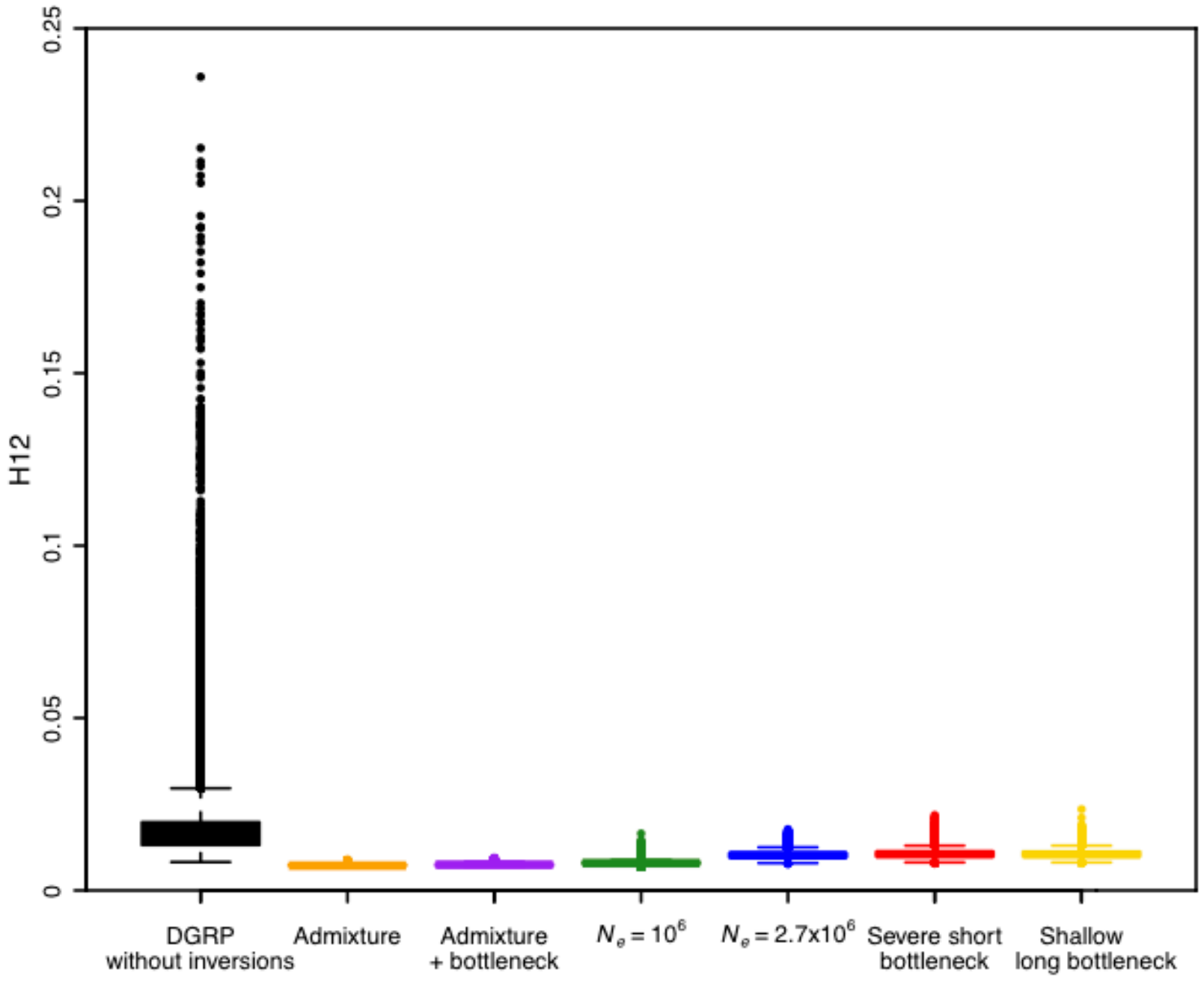

**Figure S7: H12 scan in three additional data sets of the North Carolina *D. melanogaster* population.** We reran the H12 scan in three data sets: (A) DPGP data, (B) DGRP version 2 data set, (C) the 63 DGRP version 2 strains that do not overlap the 145 strains used in the original DGRP scan. Blue and red points highlight the top 50 most extreme peaks with high H12 values relative to the median H12 value in the scan. Red points indicate peaks among the top 50 in each scan that overlap the top 50 peaks observed in the original DGRP scan. In (A), 16 peaks overlap, in (B), 40 peaks overlap, and in (C), 12 peaks overlap. Most of the overlapping peaks are among the top ranking peaks in the DGRP scan. We identify the three well-characterized cases of selection in *D. melanogaster* at *Ace*, *CHKov1*, and *Cyp6g1* in all three scans.

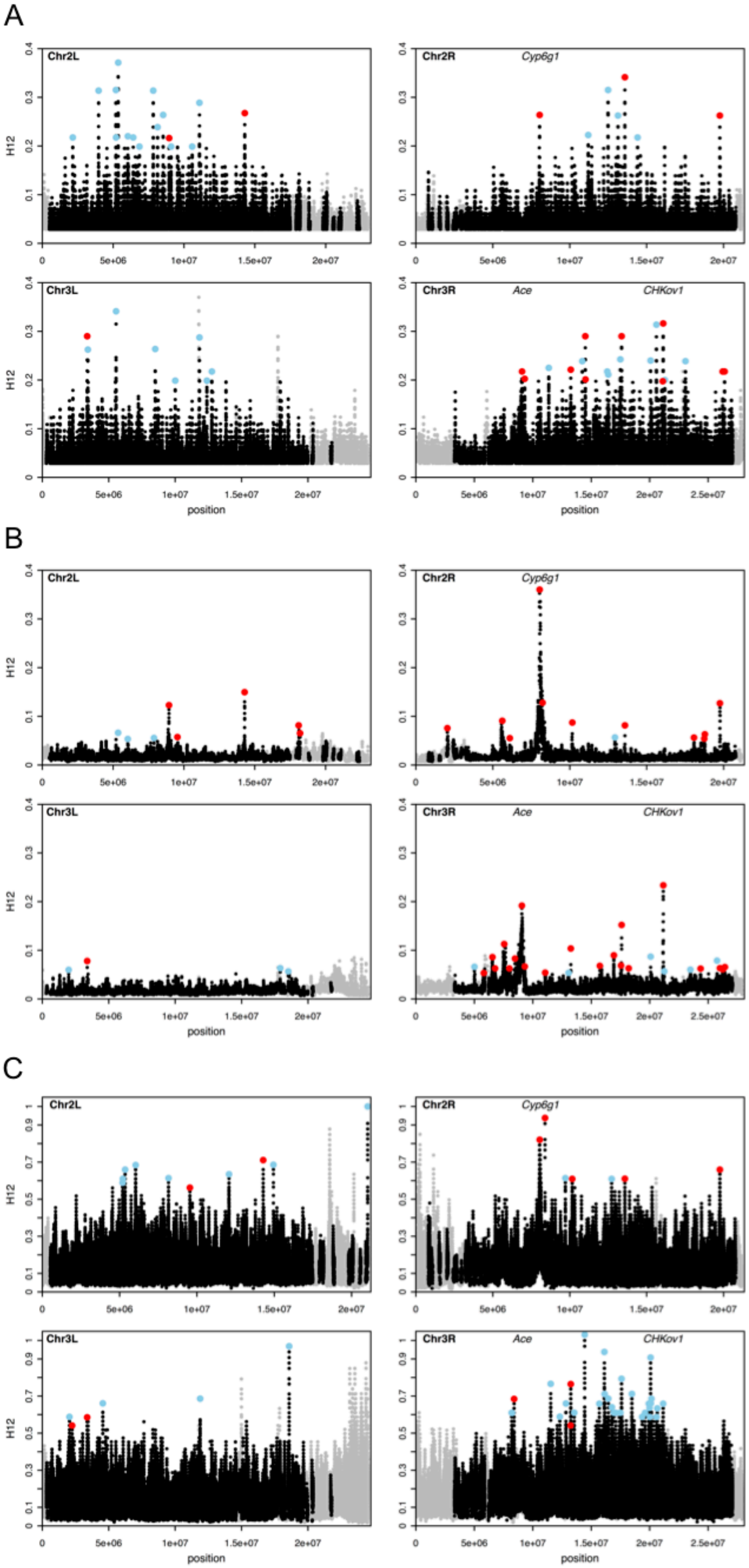

**Figure S8: Elevation in H12 values in DGRP data after down sampling to 40 strains.** DGRP strains were downsampled to 40 strains 10 times and the resulting distributions of H12 were plotted (black). In comparison to expectations under any neutral demographic model tested with a sample size of 40, all samples of 40 strains have elevated H12 values and a long tail. This indicates that the elevation of homozygosity values observed in DGRP data in Figure 7 is driven by a population-wide signal and not by any sub-population.

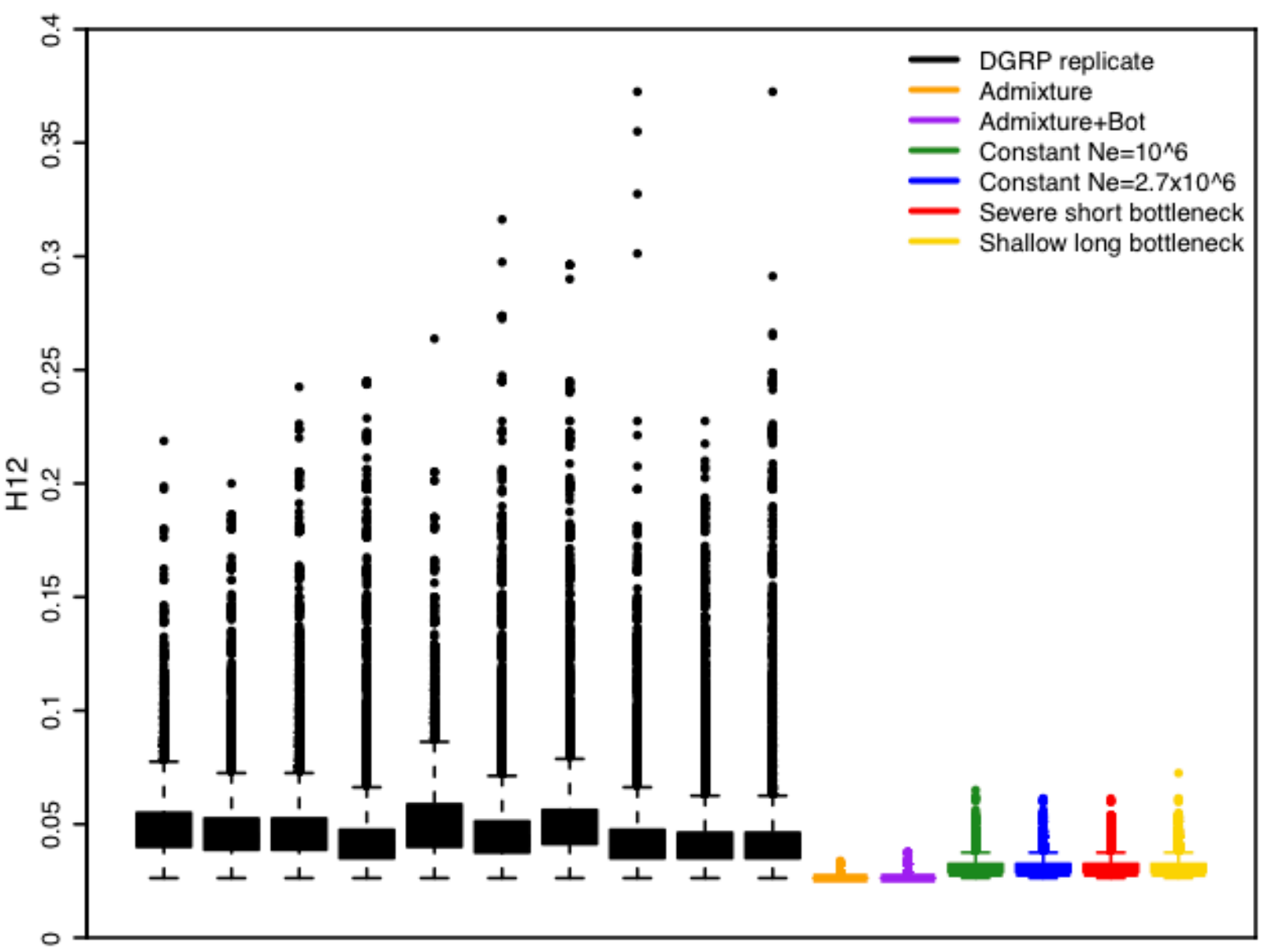

**Figure S9: H1, H12, and H123 scan of chromosome 3R.** All statistics are able to identify similar peaks. The known cases of adaptation at *Ace* and *CHKov1* have more pronounced peaks under H12 and H123.

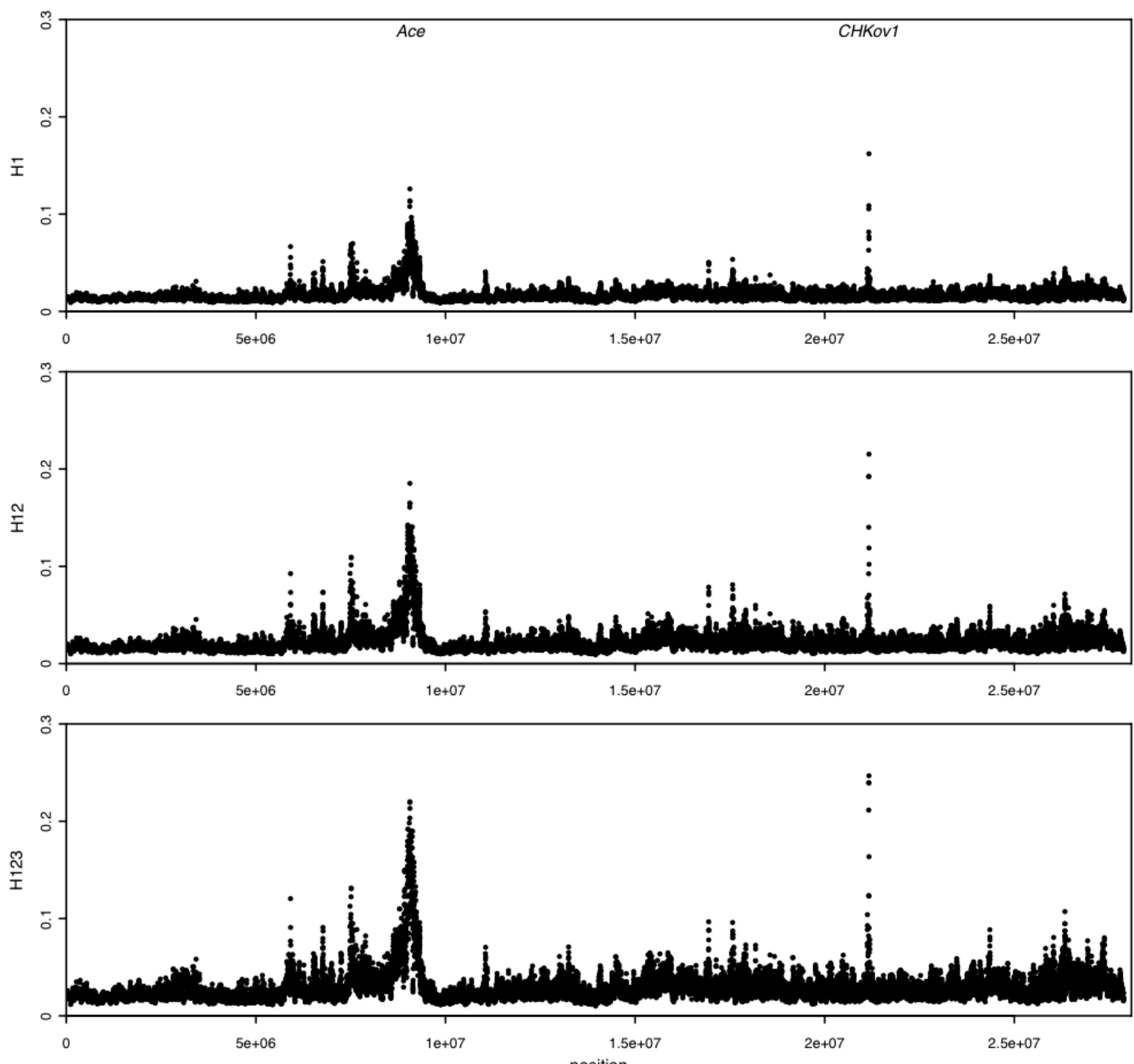

**Figure S10: Posterior distributions of $\theta_A$ and $\theta_A^{MAP}$ estimates for top peaks.** (A) Posterior distributions of $\theta_A$ measured under the constant $N_e = 10^6$ model and the admixture model (black and grey lines, respectively) and the corresponding $\theta_A^{MAP}$ estimates (dashed red and green lines, respectively) for the top nine peaks. (B) Distribution of $\theta_A^{MAP}$ values inferred under the constant $N_e = 10^6$ model for the top 50 peaks. (C) Corresponding distribution under the admixture model. The distribution of $\theta_A^{MAP}$ peaks around $\theta_A = 10$ under the constant $N_e = 10^6$ model and peaks at a slightly higher value under the admixture model, suggesting that the constant $N_e = 10^6$ model may be conservative for the purposes of inferring the softness of a sweep.

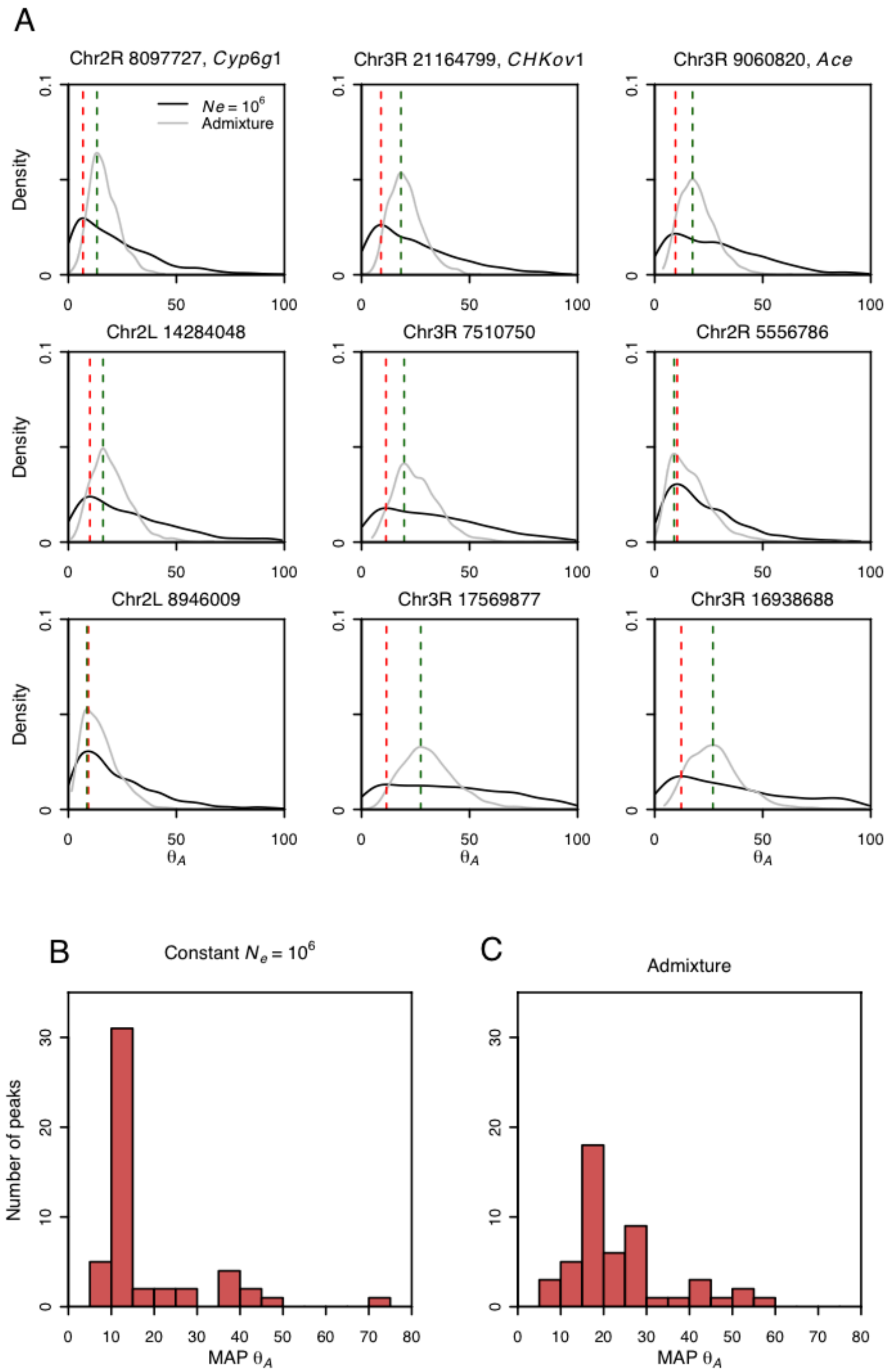

**Table S1: Parameter values used for simulations of admixture models from Figure 1.** Point estimates were calculated by Pablo Duchen (personal communication). All population sizes are in units of $N_{Ac.}$ In the admixture model (A), $N_{Ac}$=4,975,360, and in the admixture with bottleneck model (B), $N_{Aa}$=3,100,520. All times are in units $4N_{Ac}$.

**A) Admixture model**

| **Parameter** | **Symbol** | **Value** |
|---|---|---|
| Ancient size Africa | $N_{Aa}$ | 1.049994 |
| Time of bottleneck Africa | $T_A$ | 0.1192512 |
| Severity of bottleneck Africa | $sev_A$ | 0.21 |
| Current size Africa | $N_{Ac}$ | 1 |
| Time of admixture | $T_{adm}$ | 7.263e-05 |
| Proportion of European admixture | $prop_{adm}$ | 0.85 |
| Time of split Africa-Europe | $T_{AE}$ | 0.009798 |
| Ancient size North America | $N_{Na}$ | 0.0005048653 |
| Current size North America | $N_{Nc}$ | 3.2127 |
| Ancient size Europe | $N_{Ea}$ | 0.003413308 |
| Current size Europe | $N_{Ec}$ | 0.6276 |

**B) Admixture with bottleneck model**

| **Parameter** | **Symbol** | **Value** |
|---|---|---|
| Ancient size Africa | $N_{Aa}$ | 1.0401 |
| Time of bottleneck Africa | $T_A$ | 0.03241136 |
| Severity of bottleneck Africa | $sev_A$ | 0.615123 |
| Current size Africa | $N_{Ac}$ | 1 |
| Time of admixture | $T_{adm}$ | 3.757037e-05 |
| Proportion of European admixture | $prop_{adm}$ | 0.871794 |
| Time of split Africa-Europe | $T_{AE}$ | 0.006037894 |
| Current size North America | $N_{Nc}$ | 2.968357 |
| Ancient size Europe | $N_{Ea}$ | 0.004306807 |
| Current size Europe | $N_{Ec}$ | 0.7318321 |

**Table S2: Demographic parameters inferred by DaDi for simple bottleneck scenarios.** Shown are parameter estimates for six simple bottleneck scenarios fit to short intron data in DGRP inferred by DaDi and the corresponding log likelihoods for each model (LL). For all inferred models, the bottleneck sizes ($N_B$) were fixed at values as specified in the table. All population size estimates are in terms of units 4*$Ne_{\text{ancestral}}$, and all time estimates are in terms of units 2*$Ne_{\text{ancestral}}$. Values of $\theta_{\text{exp}}$ were measured for each inferred demographic models and are a function of the number of base pairs (738,024) used to generate the SFS. Note that $N_B$=0.002 represents the population size of the bottleneck inferred by Li and Stephan (2006) [55] and $N_B$=0.029 is the population size of the bottleneck inferred by Thornton and Andolfatto (2006) [34]. We ultimately chose to use the short severe bottleneck model ($N_B$=0.002, $T_B$=0.0002) and shallow long bottleneck model ($N_B$=0.4, $T_B$=0.0560) because all models fit the data equally well and these two models represent the extreme ends of the range of models tested. See Table S3 for a comparison of the fit of the severe short and shallow long bottleneck models to short intron data in terms of $S$ and $\pi$.

| $N_B$ | $N_F$ | $T_B$ | $T_F$ | **LL** | $\theta_{\text{exp}}$ |
|---|---|---|---|---|---|
| 0.002 | 0.601 | 0.0002 | 0.33 | -285.77 | 10023.95 |
| 0.029 | 0.683 | 0.0021 | 0.18 | -285.67 | 11337.58 |
| 0.05 | 0.682 | 0.0037 | 0.18 | -285.67 | 10024.82 |
| 0.1 | 0.682 | 0.0080 | 0.18 | -285.67 | 10027.46 |
| 0.2 | 0.682 | 0.0186 | 0.18 | -285.67 | 10034.67 |
| 0.4 | 0.679 | 0.0560 | 0.16 | -285.68 | 10069.48 |

**Table S3: $S$ and $\pi$ measured in neutral demographic models of North American Drosophila.** Estimates of $S$ and $\pi$ were averaged over 30,000 simulations of 10,000 bps for each demographic model. $S$ and $\pi$ estimates in DGRP short intron data were measured to be 5.8% and 1.2% per bp, respectively.

| **Demographic model** | **$S$/bp** | **$\pi$/bp** |
|---|---|---|
| Admixture | 5.8% | 1.1% |
| Admixture + bottleneck | 5.6% | 1.3% |
| Constant Ne=10^6 | 2.3% | 0.4% |
| Constant Ne=2.7x10^6 | 5.8% | 1.1% |
| Severe short bottleneck | 5.7% | 1.1% |
| Shallow long bottleneck | 5.5% | 1.1% |

**Table S4: Top 50 H12 peaks in the DGRP data.** Listed are the coordinates of the center of the analysis window with the highest H12 value in a peak, the edge coordinates of each peak, the corresponding H12 and H2/H1 values in the analysis window, the $\theta_A$ inferred for each peak and the associated 95% credible intervals for the constant $N_e = 10^6$ and admixture models, Bayes factors calculating the ratio of the likelihood of the data under a soft versus hard sweep model, and the names of the genes overlapping each peak.

| chr | center_peak | start_peak | end_peak | *H12* | *H2/H1* | theta_A_ constNe | lower95CI_ constNe | upper95CI_ constNe | theta_A_ admix | lower95CI _admix | upper95CI _admix | BF |
|---|---|---|---|---|---|---|---|---|---|---|---|---|
| Chr2R | 8097727 | 7722921 | 8486706 | 0.24 | 0.08 | 6.78 | 1.43 | 68.97 | 13.25 | 5.80 | 32.04 | 14.1538 |
| Chr3R | 21164799 | 21093966 | 21195334 | 0.22 | 0.08 | 9.01 | 1.49 | 75.40 | 18.25 | 8.24 | 39.51 | >159 |
| Chr3R | 9060820 | 8563358 | 9432483 | 0.19 | 0.13 | 9.58 | 2.00 | 76.34 | 17.50 | 7.19 | 39.34 | 124 |
| Chr2L | 14284048 | 14259362 | 14299060 | 0.12 | 0.28 | 9.98 | 1.83 | 86.17 | 16.05 | 5.93 | 41.85 | >151 |
| Chr3R | 7510750 | 7414456 | 7746544 | 0.11 | 0.25 | 11.31 | 2.86 | 87.67 | 19.72 | 9.58 | 50.99 | >244 |
| Chr2R | 5556786 | 5489549 | 5705769 | 0.10 | 0.54 | 10.39 | 2.60 | 69.69 | 8.98 | 3.79 | 43.27 | >227 |
| Chr2L | 8946009 | 8838239 | 8977912 | 0.10 | 0.50 | 9.21 | 1.82 | 74.01 | 8.61 | 3.42 | 36.10 | >22 |
| Chr3R | 17569877 | 17518429 | 17635617 | 0.08 | 0.25 | 11.51 | 3.02 | 92.73 | 27.41 | 11.45 | 64.87 | 189 |
| Chr3R | 16938688 | 16823903 | 16997106 | 0.08 | 0.31 | 12.28 | 2.26 | 91.91 | 26.95 | 9.26 | 60.02 | 104 |
| Chr2L | 18132779 | 17905347 | 18149619 | 0.08 | 0.20 | 49.28 | 4.23 | 96.45 | 43.55 | 20.79 | 88.95 | 40.7857 |
| Chr2R | 19764552 | 19741997 | 19782159 | 0.08 | 0.29 | 14.84 | 2.72 | 95.56 | 28.60 | 11.96 | 68.84 | 143 |
| Chr2R | 2043155 | 1982987 | 2232677 | 0.07 | 0.42 | 19.09 | 3.34 | 87.89 | 20.98 | 8.31 | 67.31 | 67.6667 |
| Chr3R | 6766917 | 6711552 | 6789665 | 0.07 | 0.22 | 37.91 | 5.59 | 96.65 | 40.56 | 21.31 | 91.09 | 204.667 |
| Chr3R | 26334451 | 26306832 | 26505650 | 0.07 | 0.65 | 10.72 | 3.21 | 74.38 | 9.12 | 4.47 | 45.58 | >458 |
| Chr2R | 6196252 | 6182723 | 6221969 | 0.07 | 0.53 | 10.68 | 2.22 | 84.37 | 12.12 | 5.02 | 59.71 | 130 |
| Chr2L | 18196971 | 18150953 | 18263163 | 0.07 | 0.26 | 35.80 | 5.11 | 97.63 | 43.16 | 20.44 | 91.68 | 109.833 |
| Chr3R | 7892320 | 7747647 | 7929594 | 0.06 | 0.39 | 25.11 | 3.74 | 93.89 | 25.54 | 13.71 | 82.42 | 152.6 |
| Chr3R | 26036261 | 26016366 | 26044781 | 0.06 | 0.51 | 11.68 | 2.73 | 89.76 | 15.94 | 7.04 | 75.03 | 309 |
| Chr3R | 18175477 | 18114847 | 18185495 | 0.06 | 0.50 | 14.98 | 2.26 | 87.31 | 17.45 | 7.43 | 70.57 | 375 |
| Chr2R | 18097586 | 18054633 | 18111988 | 0.06 | 0.46 | 14.99 | 2.26 | 88.00 | 19.79 | 8.73 | 66.61 | >233 |
| Chr3R | 24353929 | 24241937 | 24625912 | 0.06 | 0.58 | 12.30 | 3.53 | 88.43 | 18.68 | 7.52 | 74.51 | 592 |
| Chr2R | 5735958 | 5707191 | 5808430 | 0.06 | 0.51 | 12.69 | 3.39 | 90.38 | 20.63 | 9.87 | 79.05 | 155 |
| Chr2L | 20088273 | 20013150 | 20111508 | 0.06 | 0.42 | 20.55 | 4.39 | 93.15 | 26.09 | 13.41 | 81.42 | 498 |
| Chr3R | 17917391 | 17900909 | 17938668 | 0.06 | 0.58 | 11.80 | 2.65 | 84.85 | 14.15 | 5.85 | 76.21 | 235.5 |
| Chr2R | 18779397 | 18766062 | 18789065 | 0.05 | 0.38 | 11.65 | 3.55 | 94.27 | 26.01 | 12.73 | 77.91 | 265 |
| Chr2R | 18723092 | 18714742 | 18728140 | 0.05 | 0.54 | 10.99 | 2.62 | 89.42 | 15.30 | 7.37 | 77.53 | >244 |
| Chr2L | 9543046 | 9533192 | 9548268 | 0.05 | 0.41 | 12.99 | 2.77 | 94.01 | 19.89 | 8.98 | 70.54 | 29.1765 |
| Chr3R | 11057699 | 11027121 | 11108680 | 0.05 | 0.34 | 22.75 | 4.17 | 95.22 | 37.07 | 16.69 | 84.27 | 168.8 |
| Chr3R | 26932837 | 26901747 | 26983861 | 0.05 | 0.54 | 13.67 | 2.93 | 88.59 | 17.48 | 8.07 | 79.10 | 393 |
| Chr2R | 13587388 | 13579934 | 13591158 | 0.05 | 0.44 | 12.46 | 3.04 | 95.06 | 22.74 | 10.55 | 74.53 | >536 |
| Chr3L | 3379750 | 3355891 | 3388165 | 0.05 | 0.29 | 39.21 | 6.41 | 98.11 | 54.52 | 24.43 | 94.81 | 63.5909 |
| Chr3R | 15339462 | 15273936 | 15415377 | 0.05 | 0.55 | 12.86 | 3.25 | 88.92 | 17.47 | 8.72 | 80.46 | 87.5 |
| Chr3R | 18556910 | 18542540 | 18608604 | 0.05 | 0.27 | 42.00 | 3.87 | 97.07 | 46.24 | 18.38 | 92.56 | 15.4048 |
| Chr2L | 14851029 | 14840204 | 14892944 | 0.05 | 0.44 | 13.64 | 3.14 | 92.74 | 25.92 | 13.37 | 82.57 | 225.5 |
| Chr3R | 15864238 | 15570025 | 15993345 | 0.05 | 0.61 | 11.90 | 3.03 | 81.97 | 11.31 | 4.34 | 59.19 | >81 |
| Chr2L | 8317289 | 8293775 | 8328523 | 0.05 | 0.51 | 10.99 | 2.75 | 87.93 | 19.89 | 9.09 | 79.26 | 124.5 |
| Chr3R | 8471637 | 8333340 | 8562798 | 0.05 | 0.39 | 15.99 | 3.77 | 95.81 | 32.37 | 15.50 | 85.70 | 500 |
| Chr3R | 27035947 | 26985171 | 27253111 | 0.05 | 0.47 | 40.36 | 4.53 | 91.77 | 25.01 | 14.31 | 87.11 | 184 |
| Chr3R | 6517364 | 6496925 | 6565448 | 0.05 | 0.29 | 71.82 | 7.73 | 98.11 | 55.62 | 27.16 | 95.58 | 135.25 |
| Chr3R | 17868544 | 17842971 | 17899239 | 0.05 | 0.55 | 10.49 | 2.62 | 86.68 | 17.58 | 7.91 | 79.53 | 651 |
| Chr3R | 26272089 | 26220830 | 26304059 | 0.05 | 0.56 | 13.31 | 3.68 | 88.58 | 20.87 | 11.38 | 84.50 | >983 |
| Chr2R | 2453765 | 2331841 | 2483131 | 0.05 | 0.69 | 14.53 | 4.56 | 80.47 | 12.77 | 6.86 | 56.90 | >957 |
| Chr2R | 6101046 | 6087671 | 6125491 | 0.05 | 0.26 | 36.27 | 4.91 | 97.82 | 51.93 | 26.26 | 96.91 | 226 |
| Chr3R | 13245371 | 13195146 | 13304663 | 0.05 | 0.60 | 13.65 | 3.36 | 86.81 | 17.25 | 7.49 | 79.10 | 684 |
| Chr2R | 10140367 | 10118647 | 10172071 | 0.05 | 0.47 | 11.92 | 2.36 | 92.14 | 23.26 | 9.45 | 77.93 | 102 |
| Chr2L | 4156488 | 4148854 | 4181449 | 0.05 | 0.57 | 11.42 | 2.75 | 88.86 | 15.87 | 7.91 | 80.61 | 192.25 |
| Chr3R | 15434756 | 15416317 | 15528841 | 0.05 | 0.59 | 11.64 | 2.92 | 88.08 | 16.83 | 5.41 | 71.73 | 60.1667 |
| Chr3R | 14491226 | 14452752 | 14607381 | 0.05 | 0.52 | 11.90 | 2.65 | 90.15 | 21.63 | 9.54 | 77.53 | >983 |
| Chr3L | 2243951 | 2225398 | 2253562 | 0.05 | 0.61 | 14.44 | 3.36 | 86.81 | 16.77 | 7.41 | 77.23 | 84.3333 |
| Chr3R | 5814615 | 5761854 | 5891343 | 0.05 | 0.46 | 26.21 | 4.22 | 92.01 | 27.22 | 15.32 | 89.68 | 979.5 |

| chr | center_peak | gene_names |
|---|---|---|
| Chr2R | 8097727 | CG43190,S2P,CG34229,CG34230,Mtor,Damm,CR42532,TwdlBeta,CG42531,cuff,snoRNA:Me28S-A1322,ERp60,MCPH1,CG13189,PI31,CG33145,CG30037,CG34231,CG30036,CG8298,rho-7,RnrS,CG8964,CG8321,pds5,CG43191,128up,CG30039,Drep-1,CG13186,skpB,CG18343,CG13178,CG8407,Hen1,CG8878,mir-988,mir-281-2,mir-281-1,snoRNA:Prp8-a,CG13177,SmD3,CG34232,CG13175,CG33964,wash,CG8860,Cyp6g1,SmF,Cct5,EndoG,snoRNA:Me28S-C3420a,snoRNA:Me28S-C3420b,RpS11,CG8858,Cyp6t3,Sr-CII,CG13171,CG8854,CG13170,CG43315,CG43316,CG43244,CR43900,Vha36-2,CG8850,CG30046,CG17739,CG30203,tRNA:CR30249,tRNA:CR30250,tRNA:CR30251,garz,CG8841,CG13163,tRNA:H:48F,tRNA:HPsi,CG8490,CG8839,Den1,ana3,CR33013,CG30049,CG30043,Cpr49Aa,CG13155,Cpr49Ab,CG8501,CG13159,CG13160,Cpr49Ad,Cpr49Ac,Or49a,CG30048,CG13157,CG33627,CG33626,Cpr49Ah,CG30050,Cpr49Af,Cpr49Ag,CG43204,Nup54,CG13154,CG8520,CG8525,CG8834,CG30334,CG30051,Lac,CG8550,CG8545,CG34234,achi,vis,CG13151,CG8569,CG33632,CG33752,CG33775,CG30056,ClC-b,Ak6,stil,CR30055,CR43909,Sobp,CG13183,CG13188,Ef1alpha48D,CG13185,otk,CG8888,RpIII128,Mppe,Drep-3,Oda,Cyp6g2,CG8378,Prp8,SIP2,CG13168,Cam,CG42700,CG30047,CG34021,CG8830,Dh44-R2,Cpr49Ae,dgt5,CG42782,Dyb,fra,CG8818,Cyp301a1,Sin3A,Amph,CG8290,Sln,jeb,CG33012,fdl,s-cup, |
| Chr3R | 21164799 | CG11891,CG11889,CR13656,CG11878,CR43310,CG11892,CG31300,CG10514,CG31098,CG31104,CG13658,CG11893,CG31102,CG13659,CG31097,CG31288,CG31370,CG31436,CG10550,CG10560,CG10562,CG10553,CG10559,CG31087,CHKov2,CG10669,CG11902,tobi,rha,CG11913,CHKov1,CG31099,CG10513,Fur1, |
| Chr3R | 9060820 | CG5724,CG31345,CR33929,d-cup,CG10909,Spc25,Cyp304a1,CG14384,CG7091,Paip2,CG14383,yellow-f,CG17327,CG7488,yellow-f2,CG11656,CG8031,CG12360,l(3)87Df,CG7966,CG11668,CG11670,CG31157,Hsc70-2,snk,CG43063,CG34308,CG8138,CG8141,CG14380,CG8508,CG8483,CG8476,Ravus,CG11686,CG8449,CG15887,Osi22,wntD,CG8773,CG15888,CG32473,CG43208,CG8784,mthl12,mRpS21,CG8870,CG9813,CheA87a,CG34309,Lip3,CCHa2,CG9799,CG14374,CG33977,CG9796,yellow-e2,yellow-e3,Act87E,Ir87a,CR42756,mir-252,CG12538,CG42778,CG31337,CR43848,CG14370,CG14369,CG5999,beat-Vc,beat-Va,CG10126,beat-Vb,CtBP,CG7381,ry,pic,sim,CG31342,CG7518,timeout,2mit,CG8630,CG8774,CG8795,Ace,Su(var)3-7,CG14372,yrt,yellow-e,Dic1,CR17025,CG14377,Droj2,poly,grsm, |
| Chr2L | 14284048 | CG32971,CR43639,CR43640,wb, |
| Chr3R | 7510750 | CG6791,CG14711,CG18764,CG14710,CG6808,CG6813,Elp1,CG14715,CG18476,CG18765,Fer3,CG6908,Ho,Taf12,CG6830,CG6834,I-t,CG14717,CG14718,CG14720,mus309,CG14721,CG43062,CG6923,CG17360,HisCl1,CoVa,CG6950,glo,sad,mthl5,CG6962,CG6971,CG34307,CG14712,Jupiter,Lk6,CG12594,CG42327,CG14722,Sbf,ClC-a,CG6959,CG31368,l(3)neo38,Cad87A,Csk, |
| Chr2R | 5556786 | CG30339,CG12926,CG30000,CG30005,mir-307a,mir-307b,tRNA:M3:46A,Uba1,CG30002,CG1773,CG10459,CG1690,CheA46a,lectin-46Ca,CG1902,sqa,dap,CG1648,CG1688,CG1698,trpl,lectin-46Cb,Mmp2, |
| Chr2L | 8946009 | CG32986,CG32988,CG32987,CG32983,CG9483,CG42713,CG9510,Tsp29Fb,CG9515,Tsp29Fa,CG31886,CG32985,CG32984,CG18088,CG9541,CG13101,C1GalTA,CG9525,CG34398, |
| Chr3R | 17569877 | CG42870,CG42869,Sfp93F,CR43096,CG5849,CG31233,CG31343,CG34034,CG31198,burs,CG42335,tsl,CG6800,RpI12,mir-4969,mir-999,CG31176,GABA-B-R2,CASK, |
| Chr3R | 16938688 | CG10827,CG17278,Rlip,CG7079,CG31207,CG31189,CG12278,CG17279,Mvl,Cortactin,dmrt93B,CG7056,r-l,RhoGAP93B,rtet,CG5745,sec15,Obp93a,ppan,CG17282,slmb,CG5793,CG7009,Ubpy,CG5802,CG10824,Snmp1,Dhc93AB,CG5697,Calx,AnnIX,Ice2,Rab11,CG7044,CG5810,SNF4Agamma, |
| Chr2L | 18132779 | CR43274,CG43271,CG5681,CG31742,CG42634,CG42635,CG5693,CG31740,elfless,CG42659,ninaD,CG31741,Arr1,CG15153,CG5755,CG15152,CG31785,Socs36E,rdo,CG5674,btv,CG5758,kel, |
| Chr2R | 19764552 | TBPH,Thiolase,CG5569,CG4585,wibg,PHDP,CG4882,TM4SF,Dcp1,DNA-ligI,Upf3,bgcn,CG5597,ken, |
| Chr2R | 2043155 | tRNA:K2:42Ae,tRNA:R2:42Ad,CG14589,tRNA:N5:42Ah,tRNA:N5:42Ag,tRNA:N5:42Af,CR43904,tRNA:CR30316,tRNA:R2:42Ac,tRNA:N5:42Aa,tRNA:N5:42Ab,tRNA:K2:42Ad,tRNA:K2:42Ab,tRNA:K2:42Ac,tRNA:N5:42Ac,tRNA:K2:42Aa,tRNA:I:42A,tRNA:R2:42Ab,Cyp6w1,tRNA:R2:42Aa,tRNA:N5:42Ae,tRNA:N5:42Ad,CG8343,CG11211,CG30432,CG8335,CG30431,CG17994,l(2)k14710,Ptr,EcR,Pld,tomboy40,bin3, |
| Chr3R | 6766917 | Adk3,Tengl4,CG4674,CG6621, |
| Chr3R | 26334451 | CG31371,PH4alphaNE1,PH4alphaMP,Jon99Fi,PH4alphaSG2,Jon99Fii,mir-4908,CG31524,CG9698,PH4alphaNE2,CG15539,CG34041,PH4alphaSG1,CG34155,CG9702,Rpt6R,CG31019,CG2246,CG31016,PH4alphaNE3,CG31021,CG2267,CG31013,PH4alphaPV,CG34432,PH4alphaEFB,CG9717,tmod,jdp,CG34433, |
| Chr2R | 6196252 | CG34222,Obp46a,Ndg,CG12909,JhI-1,CAP,CG42732, |
| Chr2L | 18196971 | Socs36E,CG7200,CG15155,CG5783,CG7180,CR43413,CG31802,CG31788,CR43408,CG17681,CG43406,CG42750, |
| Chr3R | 7892320 | CG14731,CG31211,Hsp70Aa,CG3281,CG12201,CG18347,CG12213,aur,Hsp70Ab,Tango9,CG10005,CG3397,CG18547,CG12224,Tk,Ect3,CG3532,KLHL18,Spt3,CG3313,ssp5,CG31358,CG42505,CG42504,CG14739,CG14736,CG14740,mfas,dpr17,CG14741,Cad87A, |
| Chr3R | 26036261 | CG9747,snoRNA:Psi18S-1377e,snoRNA:Psi18S-1377d,snoRNA:Psi18S-1377c,snoRNA:Psi18S-1377b,snoRNA:Psi28S-2626,CG9743,snoRNA:Psi28S-2149,snoRNA:Psi18S-1377a,CG15531,CecB,Cec-Psi1,CecA2,Anp,CecA1,CecC,Cec2,RpS7, |
| Chr3R | 18175477 | mir-1010,CR43696,CR43697,CG7084,CG5386,CG33721,CG7080,CG13862,CG5391,CG5388,rdhB,CG34377,sar1,SKIP, |
| Chr2R | 18097586 | CG30279,CG3045,CG11170,CG6758,CG11275,Vps35,snoRNA:Or-CD1,CG3264,CG3290,CG3292,CG11291,CG30278,Oatp58Dc,Oatp58Da,Oatp58Db,a,MED16,ari-2,Swim, |
| Chr3R | 24353929 | CG1894,CG31051,CG12413,fkh,snoRNA:Psi28S-3305b,snoRNA:Psi28S-3405d,snoRNA:Psi28S-3405a,snoRNA:Psi28S-3305a,snoRNA:Psi28S-3405b,snoRNA:Psi28S-3405c,CG43440,Noa36,snoRNA:Psi28S-3305c,CG9986,CG31050,CG14062,CG9988,CG9989,CG33346,AR-2,CG9997,CG14061,CG12558,CG34295,Ppn,Dhc98D,MRE23,CG10011,Hrb98DE,CG10000,htt,beat-VI,CG9990, |
| Chr2R | 5735958 | dila,CG30001,CG34033,Orc6,CG1665,CG1663,CG1599,Lsm11,CG1667,CG1671,CG12744,cbx,CG30010,Ntmt,CG18446,CG12923,CG30008,CG1513,CG1441,Fmrf,CG1648,hebe,CG1516,Prosalpha7,CG12140,CG30007,Mef2,sec24, |
| Chr2L | 20088273 | CG10659,lok,vls,barr,fok,pr,CG10721,nesd,mRpS18B,Taf13,CG10747,CG13970,CG43861,bwa,Kua,CG10730,neb,sNPF, |
| Chr3R | 17917391 | CG5791,CG13407,CG5778,CG13408,how, |
| Chr2R | 18779397 | asrij,CG3499,CG3501,PIP5K59B,CG3700,MED23,nahoda,Gmer, |
| Chr2R | 18723092 | CG30265,CG12490,CG30272,CG9825,CG42284, |
| Chr2L | 9543046 | CG13113,CG13114,CG17855,Cpr30B,Oatp30B,jp, |
| Chr3R | 11057699 | CG6654,CG4203,CG4210,Spn88Eb,Spn5,CG12241,CG31344,CR43471,Caf1,Rpb7,Art3,mRpS10,CG34316,CG6499,Hsc70-4,CG42404,Su(var)3-9,Set,eIF-2gamma,CG4334,MRG15,Cp190,CG4338,l(3)neo43,CG14864,Oscp,SIDL,Tm1,CG42542,tefu, |
| Chr3R | 26932837 | CG11318,CG15553,Prosalpha3T,CG15554,CG15556,CR43458,Sox100B,CG11317,Gycbeta100B, |
| Chr2R | 13587388 | snoRNA:U27:54Eb,snoRNA:U27:54Ea,snoRNA:U29:54Ed,snoRNA:snR38:54Eb,snoRNA:U29:54Eb,snoRNA:U76:54Eb,snoRNA:U29:54Ec,snoRNA:U31:54Eb,snoRNA:U31:54Ec,snoRNA:U31:54Ed,snoRNA:snR38:54Ec,rdgBbeta,Uhg1,snoRNA:Me18S-A28a,snoRNA:Me28S-G3081a,snoRNA:Me28S-A1666a,snoRNA:Me28S-A3407a,snoRNA:Me28S-G3277a,CG6424,swi2, |
| Chr3L | 3379750 | Drs,YT521-B,CG12012,CG12014,kst,CG12010, |
| Chr3R | 15339462 | CG6255,CG15025,snoRNA:Me18S-A1374,CG31221,Dys, |
| Chr3R | 18556910 | Irp-1A,Takl2,CG17618,CG6982,CG4813,HP1c,rumi,CG6985,CG31139,CG17141,vret,CG43092,CG43091,CG13841,CG4721,CG4723,CG4725,mir-4953,CG43095,CR43654,CG43094,Dcr-1,CG7023,wge,CG43093,CG7029, |
| Chr2L | 14851029 | CG42682,CR43805,CG15279,CG15278,CG4480,Mst35Bb,Mst35Ba, |
| Chr3R | 15864238 | CR43488,CG11391,CG11453,CG11407,CG11659,tRNA:V3b:92Bb,tRNA:CR31215,tRNA:CR31471,CG31459,CG4686,CG11447,CG4572,Ire1,Pk92B,CG17186,Arc42,CG4733,Surf6,CG4465,Xport,RhoGAP92B,CG42508,mira,CR43282,CG4459,CG4462,CG4783,CG31213,CG17190,MED25,CG4433,trem,CG4424,CG4854,psidin,Indy-2,CG33934,CG4390,CG4973,Rh3,CG31206,Gr92a,CR42836,CG4662,ninaE,bnl,CG4562,Hs6st,CG4836,CG4770,CG10887,CG17193,CG6300,GluClalpha,CG4936,CG10889,CG4538, |
| Chr2L | 8317289 | Scgalpha,wol,CR43752,CG7818,CG7830,CG7806,CG7787,mtsh,CG7810,CG14275,CG7840,CG7781,Btk29A, |
| Chr3R | 8471637 | Hsp70Bb,Hsp70Bc,hug,mir-284,Vha55,CG18530,CG18616,CG11608,CG11598,CG6753,CG11600,CG6234,CG6225,CG43630,CG6188,CG14395,Cyp313a4,CG31347,Su(fu),kar,CG14394,Past1,CG12279,CG14391,Men,Octbeta3R,Snx3,mbo,mus308,Arp87C,Octbeta2R, |
| Chr3R | 27035947 | bnk,CG1544,CG15561,mir-4949,RNaseP:RNA,CG1746,CG1542,zwilch,CG15564,CG15563,mRpL32,spn-F,CG1750,CG15555,CG3669,CG18672,CG18673,CG11340,gskt,Gcn2,CG31002,CG31204,stops,Gycbeta100B,CG12054,qless,CG31004,chp,CG1607,CG11337,CG34347,Gprk2, |
| Chr3R | 6517364 | CG6465,CG31278,CG14684,CG14689,CG31373,CG14683,CG31467,pug,CG31391,CG4073,tomboy20,CG14688,Skeletor,Takr86C, |
| Chr3R | 17868544 | CG6455,Cchl,ND42,CG13409,mRpL35,CG6028,BG4,scaRNA:mgU5-38,pit,CG6015,CG6439,how, |
| Chr3R | 26272089 | CG15533,mRpS18C,CG2218,CG15536,CG15535,CG15534,Osi23,CDase,CG2224,spdo,PH4alphaEFB,CG42740,CG2217,CG15537,aralar1, |
| Chr2R | 2453765 | CR43905,jing, |
| Chr2R | 6101046 | CG12912,Hr46, |
| Chr3R | 13245371 | lute,sds22,CREG,CG43196,CG5863,CG42823,CG42798,CG17283,CG34279,CG34280,CG5860,CG42824,CG42834,CG42835,CG5866,CG33333,CG14332,CG42821,CG42822,CG14331,Eh,CG14330,Brf,Sur-8,CG5873,Dscam3, |
| Chr2R | 10140367 | mir-1016,opa1-like,CG8503,Mdr50,CG8494,beta4GalNAcTA,CG8531,CG8547,conv,Ih,CG8485,Hsc70-5,SelD, |
| Chr2L | 4156488 | CG2955,Or24a, |
| Chr3R | 15434756 | Cpr92A,CG7333,CG7342,CG17752,CG16727,CG34138,CG31220,CG31219,CG6195,Nup58,Vha13,ort,tRNA:V3b:92Ba,CG7432,CG6231,CG17751,Naam,CG6184,CG16718,Dys, |
| Chr3R | 14491226 | tRNA:CR31228,CstF-64,CG31231,CG31230,Cpsf73,Gos28,CG31229,CG7706,CG7708,CG7705,CG34282,CG14300,CG34283,CG14302,CG7715,CG7714,gwl,CG7718,CG31224,Muc91C,CG7702,VAChT,CG14301,CG14299,Mekk1,CG7720,gin,Cha, |
| Chr3L | 2243951 | osm-1,ACXD,CG9018,CG32305,CG1275,CG32301, |
| Chr3R | 5814615 | Art4,CG5359,Gr85a,Spn85F,CG3909,mtTFB2,Npc2d,Npc2e,Fancl,CG11722,CG3925,CG12811,CR33629,CG33631,CG33630,Npc2c,Mical,CG31407,Glut4EF, |

**Table S5A: Test for correlations between locations of the top 50 peaks and inversions in the DGRP data.** We performed a two-sided binomial test comparing the observed number of peaks overlapping a given inversion and the distribution of expected number of peaks overlapping an inversion. Inversions were identified by Spencer Koury (personal communication). We tested for correlations with only those inversions that were present in at least two strains. We calculated the expected number of overlapping peaks by assuming a uniform distribution of peaks throughout the genome and calculated the proportion of the genome that each inversion overlapped ('Probability of overlapping this inversion'). In all but one cases, there was no significant deviation between the observed and expected number of peaks overlapping inversions. Only for In(3R)K we found a greater than expected number of peaks overlapping the inversion. However, in Table S4B, we show that this may be due to several haplotype clusters comprised solely of two haplotypes in inversions. These haplotype clusters do not contribute to the first and second components of the sweep.

| Inversion | Number of overlapping peaks | Probability of overlapping this inversion | p-value (p-binom two sided) | Interpretation |
|---|---|---|---|---|
| all inversions | 40 | 0.879 | 0.123 | Insignificant |
| In(2L)t | 4 | 0.164 | 0.127 | Insignificant |
| In(2R)ns | 1 | 0.065 | 0.259 | Insignificant |
| In(2R)nc | 3 | 0.037 | 0.434 | Insignificant |
| In(3L)P | 1 | 0.188 | 0.001 | Lower than expected |
| In(3R)Mo | 7 | 0.163 | 0.848 | Insignificant |
| In(3R)K | 16 | 0.100 | 6.44E-06 | Greater than expected |
| In(3R)P | 11 | 0.136 | 0.096 | Insignificant |

**Table S5B: Test for correlation between haplotypes in cluster groups and haplotypes with inversions.** We performed a chi-square test to determine whether haplotypes comprising cluster groups have greater than expected number of linked inversions on the same chromosome. In this table, we report the p-values associated with this test and find that there are no significant enrichments within haplotype groups for inversions that may be linked on the same chromosome.

| Chr | Position | ChiSqVal | Df | p-value |
|---|---|---|---|---|
| Chr2R | 8097727 | 21.87 | 24 | 0.587 |
| Chr3R | 21164799 | 19.86 | 54 | 1.000 |
| Chr3R | 9060820 | 54.57 | 54 | 0.453 |
| Chr2L | 14284048 | 12.92 | 18 | 0.796 |
| Chr3R | 7510750 | 42.92 | 54 | 0.861 |
| Chr2R | 5556786 | 41.63 | 32 | 0.119 |
| Chr2L | 8946009 | 8.20 | 15 | 0.916 |
| Chr3R | 17569877 | 34.75 | 66 | 0.999 |
| Chr3R | 16938688 | 22.79 | 72 | 1.000 |
| Chr2L | 18132779 | 19.82 | 22 | 0.594 |
| Chr2R | 19764552 | 19.46 | 32 | 0.960 |
| Chr2R | 2043155 | 17.29 | 36 | 0.996 |
| Chr3R | 6766917 | 47.40 | 57 | 0.814 |
| Chr3R | 26334451 | 42.64 | 54 | 0.868 |
| Chr2R | 6196252 | 8.47 | 46 | 1.000 |
| Chr2L | 18196971 | 13.44 | 17 | 0.706 |
| Chr3R | 7892320 | 34.23 | 57 | 0.993 |
| Chr3R | 26036261 | 50.56 | 75 | 0.986 |
| Chr3R | 18175477 | 67.60 | 57 | 0.159 |
| Chr2R | 18097586 | 12.03 | 34 | 1.000 |
| Chr3R | 24353929 | 19.86 | 63 | 1.000 |
| Chr2R | 5735958 | 14.76 | 44 | 1.000 |
| Chr2L | 20088273 | 14.56 | 23 | 0.910 |
| Chr3R | 17917391 | 46.48 | 63 | 0.941 |
| Chr2R | 18779397 | 20.68 | 42 | 0.998 |
| Chr2R | 18723092 | 32.25 | 58 | 0.998 |
| Chr2L | 9543046 | 14.60 | 21 | 0.842 |
| Chr3R | 11057699 | 41.41 | 63 | 0.984 |
| Chr3R | 26932837 | 50.68 | 75 | 0.986 |
| Chr2R | 13587388 | 11.86 | 34 | 1.000 |
| Chr3L | 3379750 | 7.26 | 18 | 0.988 |
| Chr3R | 15339462 | 65.51 | 69 | 0.597 |
| Chr3R | 18556910 | 42.90 | 45 | 0.561 |
| Chr2L | 14851029 | 8.85 | 20 | 0.985 |
| Chr3R | 15864238 | 15.20 | 57 | 1.000 |
| Chr2L | 8317289 | 12.89 | 23 | 0.954 |
| Chr3R | 8471637 | 56.62 | 69 | 0.857 |
| Chr3R | 27035947 | 57.29 | 75 | 0.936 |
| Chr3R | 6517364 | 21.12 | 69 | 1.000 |
| Chr3R | 17868544 | 24.18 | 54 | 1.000 |
| Chr3R | 26272089 | 45.50 | 60 | 0.917 |
| Chr2R | 2453765 | 14.25 | 34 | 0.999 |
| Chr2R | 6101046 | 139.46 | 48 | 0.000 |
| Chr3R | 13245371 | 73.64 | 72 | 0.424 |
| Chr2R | 10140367 | 10.13 | 40 | 1.000 |
| Chr2L | 4156488 | 15.46 | 18 | 0.630 |
| Chr3R | 15434756 | 30.14 | 75 | 1.000 |
| Chr3R | 14491226 | 34.20 | 63 | 0.999 |
| Chr3L | 2243951 | 35.78 | 25 | 0.075 |
| Chr3R | 5814615 | 16.85 | 51 | 1.000 |